\documentclass[fleqn,usenatbib]{mnras}

\usepackage{newtxtext,newtxmath}
\usepackage[T1]{fontenc}

\usepackage{ae,aecompl}
\usepackage{CJK}

\usepackage{graphicx}	% Including figure files
\usepackage{amsmath}	% Advanced maths commands
\usepackage{amssymb}	% Extra maths symbols
\usepackage{multirow}
\usepackage{makecell}
\usepackage{pdflscape}
\usepackage[usenames, dvipsnames]{color}

\newcommand{\lrb}[1]{\left(#1\right)}
\newcommand{\lrbs}[1]{\left[#1\right]}

\newcommand{\angstrom}{\textup{\AA}}
\newcommand{\Msun}{$M_{\odot}$}
\begin{document}
\label{firstpage}
\pagerange{\pageref{firstpage}--\pageref{lastpage}}
\title[Faint {\it Chandra} Sources in M3]{Identifications of Faint {\it Chandra} Sources in the Globular Cluster M3}

% The list of authors, and the short list which is used in the headers.
% If you need two or more lines of authors, add an extra line using \newauthor

\author[Yue Zhao et al.]{
Yue Zhao$^{1}$\thanks{E-mail: zhao13@ualberta.ca},
Craig O. Heinke$^{1}$,
% The other people who have worked on the HST data analysis are Haldan Cohn, Phyllis Lugger, and Adrienne Cool--it would be good to include them.
Haldan N. Cohn$^{2}$,
Phyillis M. Lugger$^{2}$,
Adrienne M. Cool$^{3}$
\\
% List of institutions
$^{1}$Department of Physics, University of Alberta, CCIS 4-183, Edmonton, AB T6G 2E1, Canada\\
$^{2}$Department of Astronomy, Indiana University, 727 E. Third St., Bloomington, IN 47405, USA;\\
$^{3}$Department of Physics and Astronomy, San Francisco State University, 1600 Holloway Avenue, San Francisco, CA 94132, USA}

% These dates will be filled out by the publisher
\date{Accepted XXX. Received YYY; in original form ZZZ}

% Enter the current year, for the copyright statements etc.
\pubyear{2018}

% Don't change these lines
\maketitle
\begin{abstract}
%*****I've trimmed it to <250 words, MNRAS requirements--CH.

 We report a $30~\mathrm{ks}$  {\it Chandra} ACIS-S survey of the globular cluster M3. 
 %By analyzing a total of $\approx 30~\mathrm{ks}$ {\it Chandra} Advanced CCD Imaging Spectrometer (ACIS) observation,16 
 Sixteen X-ray sources were detected within the half-light radius ($2.3'$) with $L_X \gtrsim 2.3 \times 10^{31}~\mathrm{erg~s^{-1}}$. %The {\it Chandra} source positions combined with optical data from
  %analysed the {\it Chandra} error circles in 
  %yielding a total of 9 optical counterparts, one of which is the previously identified luminous CV corresponding to the transient super-soft source 1E1339.8+2837. 
  %of which have 
 We used {\it Hubble Space Telescope} WFC3/UVIS and ACS/WFC images to find 
 10  plausible optical/UV 
 counterparts.%, including the known CV and transient super-soft source 1E1339.8+2837. 
 We fit the spectral energy distribution of the known cataclysmic variable 1E1339.8+2837 %finding a best fit to a composite spectrum 
 with a blue ($T_\mathrm{eff} = 2.10^{+1.96}_{-0.58}\times 10^4~\mathrm{K}$, 90\% conf.) spectral component from an accretion disc, plus a red component ($T_\mathrm{eff} = 3.75_{-0.15}^{+1.05}\times 10^3~\mathrm{K}$)  potentially from a subgiant donor.
 %X-ray spectral analysis of 
 The second brightest source (CX2) has a soft blackbody-like spectrum %with $kT$ of  $0.1~\mathrm{keV}$ when fit with a neutron star atmosphere model, 
 %indicating that this source 
 suggesting a quiescent low-mass X-ray binary (qLMXB) containing a neutron star. 
 Six new counterparts %were found to 
 have obvious UV and/or blue excesses, suggesting a cataclysmic variable (CV) or background active galactic nucleus (AGN) nature. 
 Two (CX6 and CX8) have  
 proper motions indicating cluster membership, suggesting a CV nature. % and another (CX8) may also be a cluster CV. %Of the five CV candidates, 
 CX6 is blue in UV filters but red in V-I, which is 
 difficult to interpret. Two CV candidates, CX7 and CX13, show blue excesses in B-V colour but were not detected in the UV.
 %more likely to be a background AGN. 
 The other two CV candidates were only detected in the two UV bands ($\mathrm{UV_{275}}$ and $\mathrm{NUV_{336}}$), %but not in the optical bands, %leaving their nature to be verified with future deeper observations. 
 so do not have proper motion measurements, and may well be AGNs.
 One {\it Chandra} source can be confidently identified with a red straggler (a star redward of the giant branch). %in the error circle, which strongly suggests it is the true counterpart. 
 %Two sources were identified with evolved counterparts, which suggests they are members of the RS CVn subclass of chromospherically active binaries (ABs). 
 %One of the AB candidates was found near the centre of the core (with an offset $\approx 0.18 r_c$), suggesting the identification may be spurious. 
 %\green{Among the identifications, the counterpart to CX2 is likely to be a chance coincidence; CX12 and CX16 are plausible AGNs.}
 %{\color{red} We'll want to mention an estimate of how many of our associations are likely spurious here, and probably should mention how many X-ray sources are likely AGN.}
 %indicating possibilities of other source classes (e.g. a millisecond pulsar or a black hole). 
 %with 90\% confidence interval of O8-B5 or $T_\mathrm{eff} = 2.10^{+1.96}_{-0.58}\times 10^4~\mathrm{K}$, and a red component of subgiant type (M0) component, with a larger confidence interval (>90\%) of K3-M2, or $T_\mathrm{eff} = 3.75_{-0.15}^{+1.05}\times 10^3~\mathrm{K}$.}
 The observed X-ray source population of M3 appears consistent with %the expected source population, considering 
 its predicted stellar interaction rate.
 %A possible faint red counterpart to this qLMXB candidate has a large probability of a chance coincidence. 
 %One possible counterpart was found to this source with a moderate red excess. --let's see if we have strong evidence
 %Considering the low central density and high mass of M3, it is possible that this qLMXB may have been formed from a primordial binary system. {\color{red} check..}

\end{abstract}


%%%%%%%%%%%%%%%%%%%%%%%%%%%%%%%%%%%%%%%%%%%%%%%%%%

%%%%%%%%%%%%%%%%% BODY OF PAPER %%%%%%%%%%%%%%%%%%

\section{Introduction}

 Galactic globular clusters (GCs) are gravitationally-bound dense and old stellar populations harbouring $\sim 10^4-10^6$ stars. The high stellar density in the core region leads to many stellar encounters, and therefore creates a favourable environment to form a variety of close binary systems through dynamical processes \citep[e.g.][]{fabian75,hills76,camilo2005,ivanova2006}.
 %rasio2000,ivanova2005,,ivanova2010}. 
 Bright X-ray binaries ($L_X \sim 10^{36-37}~\mathrm{erg~s^{-1}}$) in globular clusters are transient or persistent low-mass X-ray binaries (LMXBs) typically harbouring accreting neutron stars (NSs) with low-mass optical companions \citep{lewin1983,grindlay1984}.
 These LMXBs are clearly produced dynamically, based on their association with the densest globular clusters in our Galaxy \citep{Clark75,Verbunt87,Verbunt03} and in other galaxies \citep{Jordan04,Sivakoff07,Jordan07,Peacock09}.
 High angular resolution X-ray observations with {\it Chandra} have revealed large numbers of faint ($L_X \sim 10^{29-34}~\mathrm{erg~s^{-1}}$) X-ray sources in globular clusters \citep{Verbunt06,Heinke2010}. 
 %Observations with X-ray facilities with an enhanced angular resolution like the {\it Chandra X-ray Observatory} has corroborated this when X-ray point sources were found to be overabundant in most GCs (e.g. \citealt{pooley2002}). These sources were found to be either luminous or faint (with ) X-ray point sources, which are thought to be associated with different close binaries (i.e. X-ray binaries or XRBs). 
 %The brightest X-ray sources turned out to be \citep{grindlay1984, lewin1983,sidoli2001}, which are revealed by their bright X-ray bursts and outbursts \citep{heinke2010}. 
 % hertz1983,verbunt1984,
 The faint X-ray population is composed of multiple source classes, including: quiescent low-mass X-ray binaries (qLMXBs) in which accretion onto the NS is thought to be stopped or at least largely suppressed \citep{campana1998,rutledge02a,heinke03d,chakrabarty14} with luminosities typically $\sim 10^4$ times fainter than during outbursts; cataclysmic variables (CVs) where white dwarfs accrete from low-mass companions \citep{hertz1983,cool1995,pooley2002,cohn10,RiveraSandoval18}; millisecond pulsars (MSPs), thought to be radio pulsars that have been spun up by accretion \citep{Bhattacharya91}, which are observed in both X-ray \citep{saito1997,Bogdanov06} and radio \citep{ransom2005,freire17}; and chromospherically active binaries (ABs) composed of two tidally-locked  non-degenerate stars, wherein fast rotation induces active coronal regions that emit (relatively faint, $L_X < 10^{31}~\mathrm{erg~s^{-1}}$)  X-rays  \citep{bailyn1990,dempsey1993,grindlay2001,heinke2005}. 
 The qLMXBs and CVs appear to be correlated with encounter rate \citep{pooley2003,heinke03d,pooley2006,Bahramian13}, while the ABs are expected to be primordial in origin \citep[][]{Bassa04,Verbunt08,Bassa08,Lu09,Huang10}.

 The globular cluster M3 (or NGC 5272) is a good target to study for several reasons. First, M3 is massive, but has a relatively low core density (central luminosity $\rho_c \approx 3.7 \times 10^3~L_\odot~\mathrm{pc^{-3}}$ according to \citealt[2010 version]{harris1996}), which %indicates a deficiency of 
 suggests a relative predominance of primordially, over 
 dynamically, formed X-ray binaries (XRBs). 
 Moreover, a less dense core facilitates the optical identification of counterparts to X-ray sources, which is difficult in denser clusters due to crowding. The X-ray sources in M3 also show a relatively dispersed distribution, which allows easier identifications and also makes them good targets for future X-ray observations with instruments that have larger collecting area but relatively larger PSFs (e.g. XMM-Newton, and the future telescope Athena). 
 Secondly, M3's position far from the Galactic Plane means it suffers relatively little reddening, making photometric studies easier and more precise. This will, in turn, support the identification of possible counterparts. %Moreover, a relatively dispersed core might indicate that most X-ray binaries might not have formed from dynamical encounters, which makes M3 an excellent target to study X-ray binaries that evolved from primordial binaries. 

%The most well-studied X-ray source in this globular cluster is the extreme soft X-ray source 1E1339 \citep{hertz1983}, which was observed to be a supersoft X-ray source with $ L_X \sim 2\times 10^{35}~\mathrm{erg~s^{-1}}$ \citep{hertz1}. The optical counterpart of this source was found to be extremely bright in the ultraviolet \citep{edmonds2004}. Besides that, detection of several faint sources were only mentioned in \citet{stacey2011} along with the study of 1E1339. 

 Previous X-ray studies of M3 have focused on the brightest X-ray source, 1E1339.8+2837 (1E1339 hereafter). %This source is the first supersoft X-ray source (SSS) observed in a globular cluster \citep{hertz1983, hertz1993}, and it has a securely identified optical counterpart \citep{edmonds2004}. 
 1E1339 was first discovered as a faint X-ray source ($L_X \sim 10^{33}~\mathrm{erg~s^{-1}}$) by the {\it Einstein Observatory} \citep{hertz1983}. It underwent a bright outburst  observed by ROSAT in 1991-1992, during which it showed a very soft spectrum ($kT \approx 20-45~\mathrm{eV}$, $L_X \sim 10^{35}~\mathrm{erg~s^{-1}}$, \citealt{hertz1993,Verbunt95}). The source returned back to quiescence with a much harder X-ray spectrum, observed by ASCA \citep{Dotani99} and {\it Chandra} ($\Gamma \sim 1.4$; see \citealt{stacey2011}, which used the same {\it Chandra} observations presented here). 
 The optical counterpart of 1E1339 was identified by \cite{edmonds2004} as a star with a very blue $U-V$ colour, showing marked variability on timescales of hours. 
 
 1E1339 is the only supersoft X-ray source (SSS) identified in a Galactic GC, though three transient SSSs have been identified in M31 GCs \citep{Henze09,Henze13}, two of them clearly identified with nova explosions, which are the most frequent class of transient SSSs in M31 \citep[e.g.][]{Henze11}.  The bright outburst and the supersoft spectrum suggest a physical connection to other galactic supersoft X-ray sources (SSSs). However, 1E1339's peak observed X-ray luminosity was $\sim 100$ times fainter than that of standard SSSs, suggesting a much smaller burning area.

 The present work focuses on a systematic multiwavelength study of the faint X-ray sources in M3. The paper is organized as follows: in section \ref{sec2}, we report the {\it Chandra} and {\it HST} data we used in our studies. In section \ref{sec3}, we describe our analyses including data reduction, source detection, and the relevant techniques and methodologies used in astrometry, photometry, and counterpart identifications. In section \ref{sec4}, we discuss the possible nature of each X-ray source based on its photometric and spectroscopic properties. Finally, in section \ref{sec5}, we summarise our results. %and present a conclusion.

\section{Observations}
\label{sec2}
\subsection{{\it Chandra} Observations}
\label{sec:sec2.1}

 The globular cluster M3 was observed by the Advanced CCD Imaging Spectrometer (ACIS-S) on board the {\it Chandra X-ray Observatory}, using the {\it very faint} mode. Three observations, at roughly 6-month intervals, were taken, focused on monitoring 1E1339. A 1/2 subarray was used to reduce the frametime, and thus pileup from this relatively bright source, but the field of view still covers the whole half-light region of the cluster. Observation details are listed in Tab. \ref{tab:chandra_obs}.

\begin{table}
	\centering
	\caption{{\it Chandra} Observations}
	\label{tab:chandra_obs}
	\begin{tabular}{cccc}
		\hline
		ObsID & Time of observation & Exposure time $\lrb{\mathrm{ks}}$ & Chip \\
		\hline
		$4542$ & 2003-11-11 16:33:18 & $9.93$ & ACIS-S \\
		$4543$ & 2004-05-09 17:26:32 & $10.15$ & ACIS-S \\
		$4544$ & 2005-01-10 08:54:31 & $9.44$ & ACIS-S\\
		\hline
	\end{tabular}
\end{table}

\begin{figure*}
    \centering
    \includegraphics[scale=0.58]{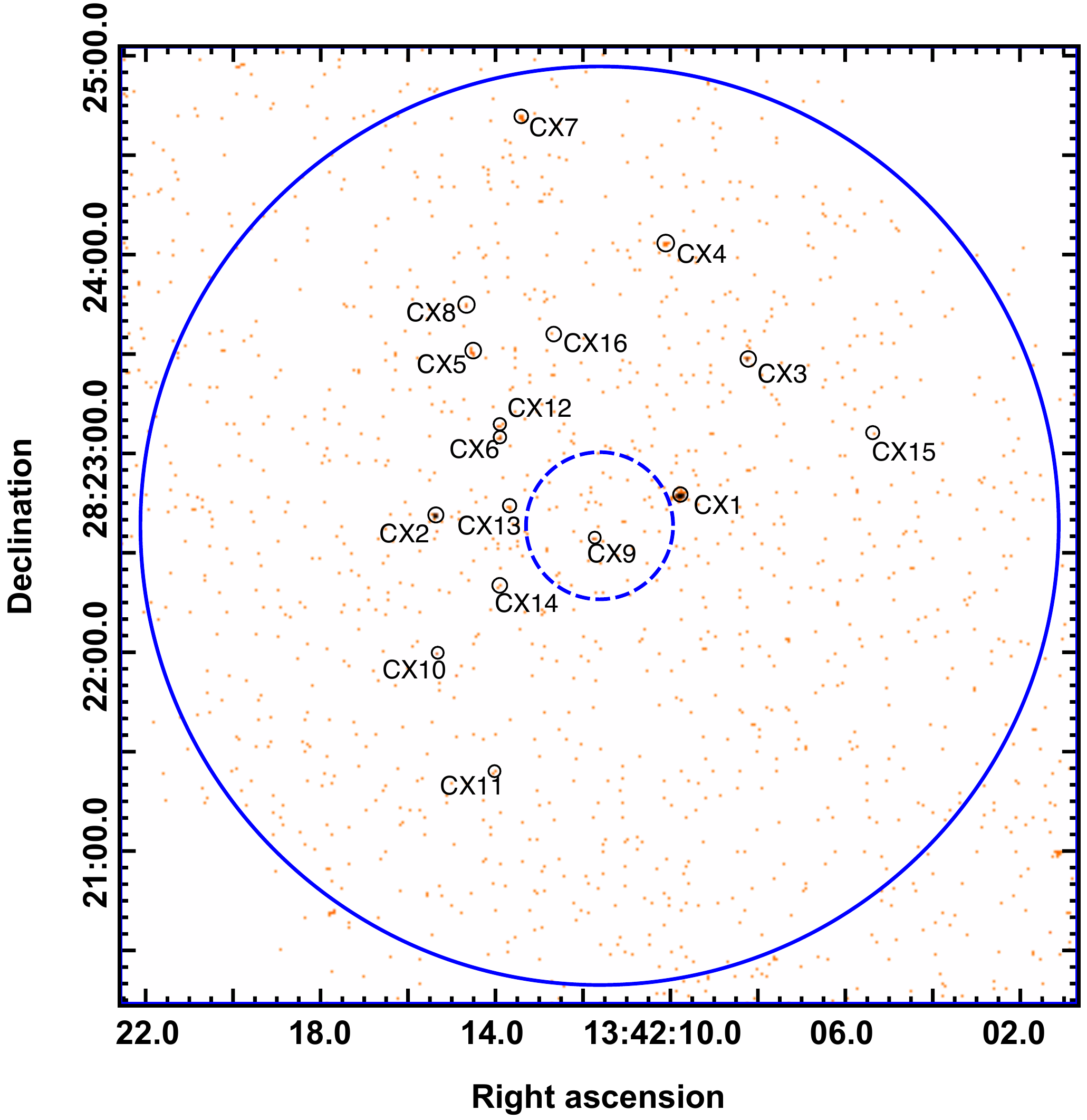}
    \caption{Merged $0.5-7~\mathrm{keV}$ {\it Chandra} exposure-corrected  ACIS-S X-ray image of the central $4{\farcm}8 \times 4{\farcm}8$ region of M3. The dashed blue circle shows the $0{\farcm}37$ core region; the solid blue circle shows the $2{\farcm}31$ half-light radius \citep[2010 edition]{harris1996}. Sources detected by {\it wavdetect} are marked with solid black circles. The size of each pixel is $0{\farcs}492$.}
    \label{fig1}
\end{figure*}

\subsection{{\it HST} Observations}
\label{sec:sec2.2}

We used {\it HST} WFC3/UVIS (GO12605, PI: Piotto), ACS/WFC (GO10775, PI: Sarajedini) and ACS/HRC and ACS/WFC (GO10008, PI; Grindlay) imaging data to search for possible optical counterparts. The WFC3 2012 data contained observations in the F275W ($\sim \mathrm{UV}$), F336W ($\sim \mathrm{NUV}$), and F438W  ($\sim \mathrm{B}$) filters, while the 2006 ACS data used the F606W ($\sim \mathrm{V}$) and F814W ($\sim \mathrm{I}$) filters.
%The WFC3 data set taken on May 15, 2012 contains a 2.49 ks exposure in the F275W ($\sim \mathrm{UV}$) filter, a 1.4 ks exposure in the F336W ($\sim \mathrm{NUV}$) filter, and a 168 s exposure in the F438W  ($\sim \mathrm{B}$) filter. The ACS data set was taken on July 1, 2006, with a 532s exposure in the F606W ($\sim \mathrm{V}$) filter and a 612s exposure in the F814W ($\sim \mathrm{I}$) filter. 
The ACS/HRC and ACS/WFC 2004 data were taken over a broad range of UV and optical bands, in conjunction with the second {\it Chandra} observation. This enables us 
to construct a simultaneous spectral energy distribution (SED) for the supersoft X-ray source 1E1339. Details of these observations are listed in Tab. \ref{tab:hst_obs}.

\begin{table*}
    \centering
    \caption{{\it HST} Observations}
    \begin{tabular}{ccccc}
    \hline
    \hline
    GO                          & Time of observation & Exposure time $\lrb{\mathrm{s}}$ & Camera/Channel & Filter\\
    \hline
    \multirow{6}{*}{10008}      & 2004-05-09 17:51:18 & $620$  & ACS/HRC   & F220W\\
                                & 2004-05-09 17:41:19 & $464$  & ACS/HRC   & F250W\\
                                & 2004-05-09 17:17:33 & $1200$ & ACS/HRC   & F330W\\
                                & 2004-05-09 19:05:02 & $680$  & ACS/WFC   & F435W\\
                                & 2004-05-09 19:22:02 & $678$  & ACS/WFC   & F555W\\
                                & 2004-05-09 18:52:26 & $290$  & ACS/WFC   & F814W\\
    \hline
    \multirow{2}{*}{10775}      & 2006-02-20 11:16:31 & $532$  & ACS/WFC   & F606W ($\sim \mathrm{V}$)\\
                                & 2006-02-20 12:55:54 & $612$  & ACS/WFC   & F814W ($\sim \mathrm{I}$)\\
    \hline
                                & 2012-05-15 01:39:53 & $2490$ & WFC3/UVIS & F275W ($\sim \mathrm{UV}$) \\
    12605                       & 2012-05-15 01:49:19 & $1400$ & WFC3/UVIS & F336W ($\sim \mathrm{NUV}$) \\
                                & 2012-05-15 01:36:38 & $168$  & WFC3/UVIS & F438W ($\sim \mathrm{B}$)\\
    
    \hline
    \end{tabular}
   
    \label{tab:hst_obs}
\end{table*}

\section{Analyses}
\label{sec3}
\subsection{Merging Chandra Observations}
\label{sec:sec3.1}
 The {\it Chandra} data were reduced using the {\it Chandra Interactive Analysis of Observations} software (CIAO) version 4.10\footnote{\url{http://cxc.harvard.edu/ciao/}} and CALDB version 4.7.8. We first reprocessed the data using the CIAO {\it chandra\_repro} script to update the calibration, generating new level-2 event files. We then add the three observations to get a deeper view of the cluster. We first adjust aspect solutions between observations by using the brightest source 1E1339. The new aspect solutions were applied to the event files by using the CIAO {\it wcs\_update} tool. Finally, the combined exposure map and the corresponding exposure-corrected image were generated by using the CIAO {\it merge\_obs} tool.  The resulting merged broad-band exposure-corrected ($0.5-7~\mathrm{keV}$) X-ray image is shown in Fig. \ref{fig1}.

\subsection{Source Detection}
\label{sec:sec3.2}

 We generated an X-ray source list by running the CIAO {\it wavdetect}\footnote{\url{http://cxc.harvard.edu/ciao/threads/wavdetect/}} tool on the combined broad-band X-ray image. The {\it wavdetect} algorithm correlates possible source pixels with a ``Mexican Hat'' function with different scale sizes and identifies pixels with sufficiently large positive correlation values to further calculate source positions, error circles and other information about the sources. We chose ${\it scales}= 2, 4$ to cover the possible sizes of point sources at different off-axis angles, and used ${\it sigthresh}=3\times 10^{-6}$ (the reciprocal of the area of the region) to limit the expected number of false detections to $1$. The detected sources are listed in Tab. \ref{tab:chandra_catalog}. The right ascensions and declinations are the coordinates as calculated by {\it wavdetect}. $P_\mathrm{err}$ is the $95\%$ error circle following the empirical formula from \citet{hong2005}. 
 
 \begin{table*}
    \centering
    \caption{A catalogue of X-ray sources in M3.}
    \begin{tabular}{ccccccccc}
    \hline
    \hline
             &  \multicolumn{2}{c}{Positions$^a$} & & & \multicolumn{3}{c}{Net Counts (absorbed)} & \\
        Name & $\alpha$ (J2000)  & $\delta$ (J2000) & Dist.$^b$ &$\mathrm{P_\mathrm{err}}^c$ & $0.5-7.0~\mathrm{keV}$   & $0.5-2.0~\mathrm{keV}$ & $2.0-7.0~\mathrm{keV}$ & Notes \\
             & $\lrb{\mathrm{h:m:s~}}$ & $\lrb{\mathrm{\deg:\arcmin:\arcsec}}$ & ($\arcsec$) & $\lrb{\arcsec}$ & \multicolumn{3}{c}{$\mathrm{counts~(err)^d}$} & \\
    \hline
        CX1    &  13:42:09.771    &  +28:22:47.618  & 25.8 &  0.295 & 1038.1 (32.3)& 738.6(27.2) & 299.5 (17.4)& SSS (1E1339)\\
        CX2    &  13:42:15.364    &  +28:22:41.458  & 49.5 & 0.321 & 138.4 (11.8) & 137.5 (11.7) & 1.0 (1.0)& qLMXB\\
        CX3    &  13:42:08:218    &  +28:23:28.528  & 67.5 & 0.362 & 41.8 (6.5)& 23.9 (4.9) & 17.8 (4.2)& -\\
        CX4    &  13:42:10.105    &  +28:24:03:498  & 87.6 & 0.410 & 19.8 (4.4)& 10.0 (3.2) & 9.8 (3.2)& -\\
        CX5    &  13:42:14.511    &  +28:23:31.053  & 64.4 & 0.382 & 18.6 (4.5)& 13.8 (3.7) & 4.8 (2.2)& -\\
        CX6    &  13:42:13.899    &  +28:23:04.927  & 40.2 & 0.403 & 16.0 (4.0)& 3.0 (1.7) & 13.0 (3.6)& CV\\
        CX7    &  13:42:13.411    &  +28:24:41.737  & 125.8& 0.519 & 8.8 (3.0)& 5.0 (2.2) & 3.9 (2.0)& CV/AGN\\
        CX8    &  13:42:14.664    &  +28:23:44.921  & 78.7 & 0.482 & 7.9 (2.5)& 4.9 (2.2) & 2.9 (1.7)& CV?\\
        CX9    &  13:42:11.730    &  +28:22:34.608  & 3.9  & 0.490 & 7.8 (2.8)& 5.0 (2.2) & 3.0 (1.7)& RS/AB\\
        CX10   &  13:42:15.322    &  +28:21:59.867  & 62.1 & 0.539 & 7.0 (2.8)& 3.0 (1.7) & 4.0 (2.0)&-\\
        CX11   &  13:42:14.023    &  +28:21:24.128  & 80.6 & 0.620 & 6.0 (2.6)& 5.0 (2.2) & 1.0 (1.0)&-\\
        CX12   &  13:42:13.901    &  +28:23:08.819  & 42.9 & 0.498 & 5.8 (2.5)& 6.0 (2.4) & 0.0 (0.0)& CV/AGN\\
        CX13   &  13:42:13.678    &  +28:22:44.243  & 27.8 & 0.453 & 5.8 (2.4)& 4.9 (2.2) & 1.0 (1.0)& CV/AGN\\
        CX14   &  13:42:13.903    &  +28:22:20.201  & 35.1 & 0.543 & 4.9 (2.2)& 3.9 (2.0) & 1.0 (1.0)& AB?\\
        CX15   &  13:42:05.389    &  +28:23:06.513  & 87.0 & 0.784 & 4.0 (2.0)& 2.0 (1.4) & 2.0 (1.4)& -\\
        CX16   &  13:42:12.668    &  +28:23:36.034  & 59.4 & 0.534 & 3.9 (2.0)& 3.0 (1.7) & 1.0 (1.0)& CV/AGN\\
    \hline
    \multicolumn{3}{l}{$^a$ Coordinates from {\it wavdetect}}\\
    \multicolumn{3}{l}{$^b$ Offsets from the center of the core in arcsec}\\
    \multicolumn{3}{l}{$^c$ 95\% error circles as calculated by \citet{hong2005}}\\
    \multicolumn{3}{l}{$^d$ errors generated by CIAO {\it srcflux} tool}\\
    \end{tabular}
    \label{tab:chandra_catalog}
\end{table*}

\subsection{Source counts}
\label{sec:sec3.3}
 We used the CIAO {\it srcflux}\footnote{\url{http://cxc.harvard.edu/ciao/threads/fluxes/}} script to calculate the source counts in 3 energy bands: broad ($0.5-7.0~\mathrm{keV}$), soft ($0.5-2.0~\mathrm{keV}$), and hard ($2.0-7.0~\mathrm{keV}$). The effective energy of each band was calculated as the flux-weighted average using the best-fit models of 1E1339 in \citet{stacey2011}. To calculate the combined counts, we first apply the script to each individual observation, and then add up the counts. For all sources except CX6 and CX12, the extraction region is defined by a circle that encloses $90\%$ of the PSF at $1~\mathrm{keV}$, and the background region is an annulus with inner radius the same as that of the extraction region and outer radius $5$ times the radius of the source region. Because CX6 and CX12 are close to one another, their background regions were defined separately as annuli excluding the other source. In Tab. \ref{tab:chandra_catalog}, we ordered the sources in descending order of their observed counts in the broad band.

\subsection{Optical Photometry}
We used the wide-field observations from three epochs (2004, 2006, and 2012) to systematically study the photometry of this GC.

The 2012 WFC3/UVIS photometry has been analysed as part of the {\it Hubble Space Telescope UV Legacy Survey of Galactic Globular Clusters} \citep[GO-13297]{piotto2015}, and the reduced data and data products, including magnitudes of detected point sources, are now available to the public \citep{soto2017}. However, we found that some stars detected on our images, which coincide with X-ray error circles, are not included in the Padova catalogue, though evidence for their presence is visible to the eye. Therefore, we also generated our own photometric catalogue by performing PSF-fitting photometry on the three WFC3/UVIS filters for counterpart searching. %To obtain further photometric information about the candidate counterparts, we also incorporated the 
We used {\it HST} data products that are pipeline-processed, flat-fielded, and with charge transfer efficiency (CTE) trails removed (i.e., FLC images) to get the stacked images  for photometry, absolute astrometry, and counterpart searches. We used the {\it HST} Drizzlepac software version 2.0\footnote{\url{http://www.stsci.edu/hst/HST_overview/drizzlepac}} for image alignment and combination. For each filter, all FLC frames are first aligned to a reference frame (one of the FLC images) with the Drizzlepac {\it Tweakreg} tool. The aligned images were then combined using the Drizzlepac {\it AstroDrizzle} tool with {\it pixfrac}=1.0. In order to get higher resolution, the drizzle-combined images were oversampled by a factor of two so that their pixel size is half of the original ($0\farcs02/\mathrm{pixel}$ for WFC3/UVIS, and $0\farcs025/\mathrm{pixel}$ for ACS/WFC). The drizzle-combined images of ACS/WFC were used for visual inspection of candidates and for making finding charts.

We then performed aperture and PSF-fitting photometry on the drizzle-combined WFC3 images (2012 observation) using the PyRAF\footnote{\url{http://www.stsci.edu/institute/software_hardware/pyraf}} DAOPHOT \citep{stetson1987} package. A star list was first generated for each filter by the {\it daofind} task with $3$ $\sigma$ detection threshold. We then chose relatively bright and isolated stars to model the PSFs. In order to account for possible spatial variability, at least $100$ PSF candidate stars were selected across each image and the PSF model was set to be a second-order function of $x$ and $y$ ($varorder = 2$). The best-fit PSF model was then applied to all stars in the field by using {\it allstar}.

The following three steps were taken to calibrate the DAOPHOT/{\it allstar} photometry. First, magnitudes corresponding to a finite aperture were calibrated to those corresponding to an infinite aperture using the "curve of growth" method applied to a subset of reasonably isolated bright stars. Second, instrumental magnitudes of infinite aperture were shifted to the VEGAMAG system by using the WFC3 zeropoints from the STScI web page\footnote{\url{http://www.stsci.edu/hst/wfc3/phot_zp_lbn}}. Finally, PSF-fitting photometry only gives relative magnitudes, so we cross-identified stars between aperture photometry and PSF-fitting photometry, and applied the average offsets to convert magnitudes in PSF photometry to instrumental magnitudes.

%{\color{red} The bold text confuses me.  What are PSF magnitudes? What are aperture magnitudes? Where do these lists come from?}

The photometry of the 2006 ACS/WFC observations (GO-10775, PI: Sarajedini) has been produced as part of the {\it ACS Globular Cluster Treasury Program} \citep[GO-10775]{sarajedini2007,anderson2008} which provides $\mathrm{V_{606}}$ and $\mathrm{I_{814}}$ magnitudes (available at the {\it Mikulski Archive for Space Telescopes} (MAST) website\footnote{\url{https://archive.stsci.edu/prepds/acsggct/}}). 

We obtained photometry for the 2004 F435W (B) and F555W (V) images (GO-10008, PI: Grindlay) using software based on the program developed for the {\it ACS Globular Cluster Treasury Program}, described in \cite{anderson2008} and known as Ksync. We performed photometric calibration to the VEGAMAG system by doing aperture photometry on moderately bright, isolated stars within a 0.15 arcsec radius aperture, finding the aperture correction to an infinite radius aperture from \citet{sirianni2005}, calculating the median offset between the Ksync photometry and the aperture photometry, and applying the calibrations from the HST calibration website.\footnote{https://acszeropoints.stsci.edu/}. 

\begin{figure*}
    \centering
    \includegraphics[scale=0.5]{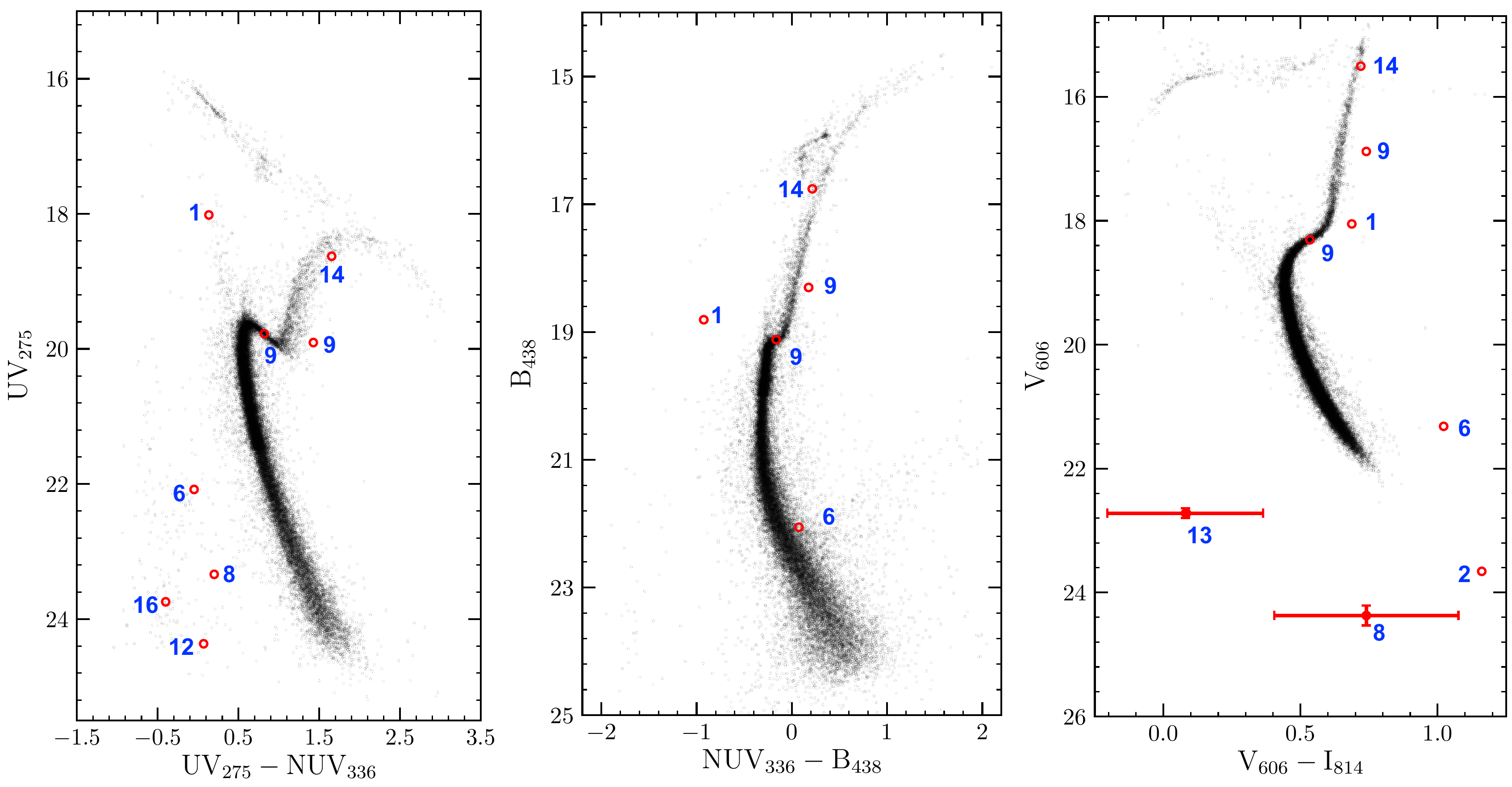}
    \caption{{\it Left:} $\mathrm{UV_{275}-NUV_{336}}$ CMD generated by stars within a region of $140\arcsec \times 140\arcsec$ centered on the cluster core. {\it Middle:} $UV_{336}-B_{438}$ CMD generated by stars within the same region. Both left and middle panels were generated using DAOPHOT results. {\it Right:} $\mathrm{V_{606}-I_{814}}$ CMD from the catalogue of the {\it ACS Globular Cluster Treasury Program}. Optical counterparts are shown with red circles and annotated with their CX IDs (two potential counterparts are shown for CX9, though we find the brighter, redder one more likely). The error bars of CX8 and CX13 were indicated to show the uncertainty of their blue excesses.}
    \label{fig:fuvcmd}
\end{figure*}

\begin{figure*}
    \centering
    \includegraphics[scale=0.5]{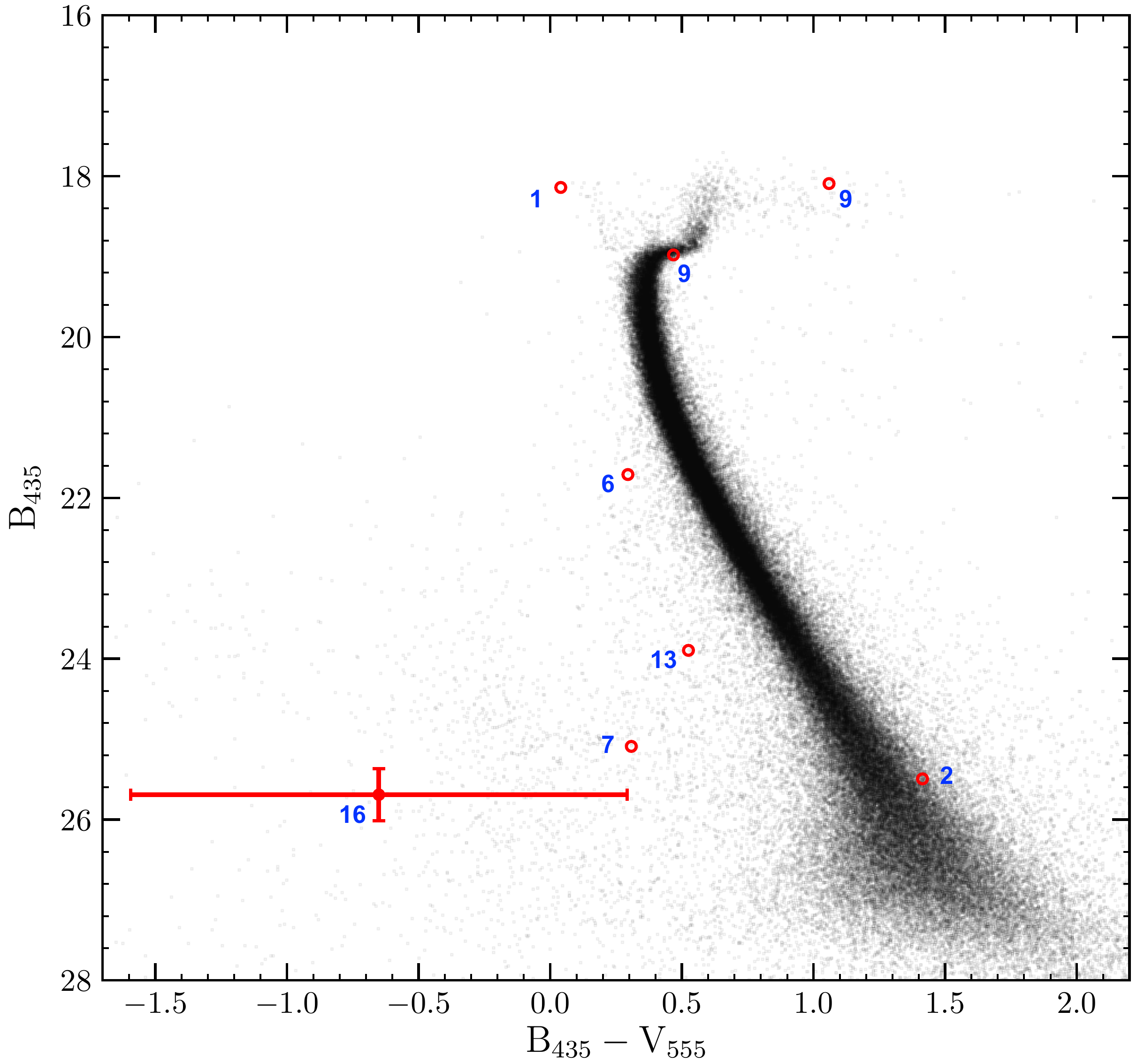}
    \caption{$\mathrm{B_{435}-V_{555}}$ CMD from the 2004 ACS/WFC observation. Optical counterparts are annotated by red circles. The large photometric errors of CX16 are indicated with error bars.}
    \label{fig:ksync_cmd}
\end{figure*}

With the obtained magnitudes for different filters, we constructed colour-magnitude diagrams (CMDs) by cross-matching catalogues in different filters. In Fig. \ref{fig:fuvcmd} we show the resulting $\mathrm{UV_{275}-NUV_{336}}$, the $\mathrm{NUV_{336}-B_{438}}$, and the $\mathrm{V_{606} - I_{814}}$ CMDs; in Fig. \ref{fig:ksync_cmd}, we show the $\mathrm{B_{435}-V_{555}}$ CMD.

\subsection{Astrometry}
 The accuracy of {\it HST} astrometry is limited by at least two error sources. The first type of error comes from the fact that positions of guide stars used to derive the astrometric information are known with some uncertainties ($\lesssim 200~\mathrm{mas}$ for WFC3; $\lesssim 300~\mathrm{mas}$ for ACS. See ACS and WFC3 Handbooks; \citealt{lucas2016}, \citealt{deustua2016}). Secondly, errors are also introduced when the instrument aperture is mapped to the guide stars. The {\it Chandra} images are not astrometrized at the sub-arcsecond level, and therefore also require astrometric calibrations.

  Our process of calibrating for astrometry includes two steps: 1. Calibrate one of the optical images to a known catalogue that has superior astrometry, and use this image as the reference frame; 2. Align other optical images to the reference frame, and align the {\it Chandra} catalogue to the reference frame for boresight correction--such that CX1 centres on its known counterpart \citep{edmonds2004}. To fulfill this, we cross-matched source positions in one of our {\it daofind} catalogues with those from the Gaia Catalogue (Data Release 2; see \citealt{gaia2016a,gaia2018_description}). We chose the $\mathrm{NUV_{336}}$ image as the reference image because it has a sufficiently long exposure, yet not so long to render bright sources saturated. We used a total of 895 Gaia sources that have relatively accurate positions (with errors in RA and DEC both $\leq 0.5~\mathrm{mas}$) found within a $1\arcmin$ search radius centred on the cluster core. Cross-matching resolved a total of 868 matches. We calculated the average offsets ($\mathrm{Gaia-NUV_{336}}$) in RA and DEC, finding an average $\Delta\mathrm{RA}\sim -0.384\arcsec \pm 0.004\arcsec$ and an average $\Delta \mathrm{DEC}\sim 0.254\arcsec \pm 0.003\arcsec$  ($1$ $\sigma$ errors). For boresight correction, we found the $\mathrm{Gaia} - Chandra$ offsets of CX1 to be $\Delta \mathrm{RA} \sim -0.22\arcsec$ and $\Delta \mathrm{DEC} \sim 0.10 \arcsec$.

\subsection{Counterpart search}
\label{subsec:counterpart_search}
 The principle of hunting for optical counterparts to XRBs more or less depends on the source class. For example, CVs typically have a strong UV excess that originates from the shock-heated region on the WD surface and/or the accretion disc, so they usually appear as blue outliers on $\mathrm{UV-NUV}$ or $\mathrm{NUV-B}$ CMDs. ABs are either K/M type main sequence stars (BY Draconis, or BY Dra) or F/K type subgiants (RS Canum Venaticorum or RS CVn). Examples of works using these classifications include \citet{cohn10} and \citet{Cool13}. Using these criteria, we searched possible counterparts primarily in the 95\% error circles. %However, considering very faint sources ($\lesssim 10~\mathrm{counts}$) might have worse localizations, 
 If no interesting object was found within the 95\% error circle,
 we also applied a somewhat larger searching region ($\leq 1.8 P_\mathrm{err}$), recognising that this procedure incurs a higher risk of spurious coincidences.
 
 We also have to consider possible confusion from foreground stars and background AGNs. %These objects usually disguise as optical counterparts in the sense that 
 Foreground stars usually appear redder, while background AGNs appear bluer, than cluster members. To exclude non-members, we check the proper motion (PM) of each counterpart candidate and compare that with the PM of the GC. The cluster membership can be confirmed if the star moves in accordance with the cluster's systematic PM, and in disagreement with the PM of other possibilities. The enhanced angular resolution of HST cameras (e.g. ACS and WFC3) has made proper motion measurement possible over a relatively short span of time. Therefore, as part of the {\it Hubble Space Telescope UV Legacy Survey of Galactic Globular Clusters}, \cite{piotto2015} includes a PM study of M3 by cross-matching the WFC3 source list with the ACS source list obtained 6.2 years earlier. Using the ACS(2006) and WFC3(2012) $x$ and $y$ positions from the Padova catalogue, we first calculated the displacements in $x$ and $y$ for each counterpart. The displacement can then be converted to proper motions with $v_i = \Delta x_i\times 50/\mathrm{Epoch}$ (see \citealt{soto2017}), where $\mathrm{Epoch} = 6.2~\mathrm{yr}$ is the time interval between the ACS and the WFC3 observations. As another check on cluster membership, we incorporated the membership probability ($P_\mu$) from the 2018 release of the public catalogue (see \citealt{nardiello2018}). 
 
 We used stars that have at least one good photometric measurement in all three UV filters from the Padova catalogue as the sample for our PM check. We first computed PMs for each candidate counterpart and then compare them with the PM rms of the sample. A star is accepted as a cluster member if both its $v_\alpha$ and $v_\delta$ are smaller than $3$ times the PM rms ($\approx 1.779~\mathrm{mas~yr^{-1}}$).  Fig. \ref{fig:vpd} shows the PM distribution of the selected sample plotted in 4 separate magnitude bins, with candidate counterparts marked with red circles. The 3-rms limit of each bin is indicated with a dashed circle. Because the first bin is made of bright stars that are close to saturation, it has a relatively large rms.
 %{\color{red} maybe add a circle for plausible AGN, at 3 sigma, too.} 
 To exclude background AGN, we used the proper motion measurement of the cluster from GAIA DR2 \citep[see Table C.1.]{gaia2018}, from which we found $v_\alpha = 0.1127 \pm 0.0029~\mathrm{mas~yr^{-1}}$ and $v_\delta = 2.6274\pm 0.0022~\mathrm{mas~yr^{-1}}$ for background AGN (marked with a red cross in Fig. \ref{fig:vpd}).
 
\begin{figure}
     \centering
     \includegraphics[scale=0.56]{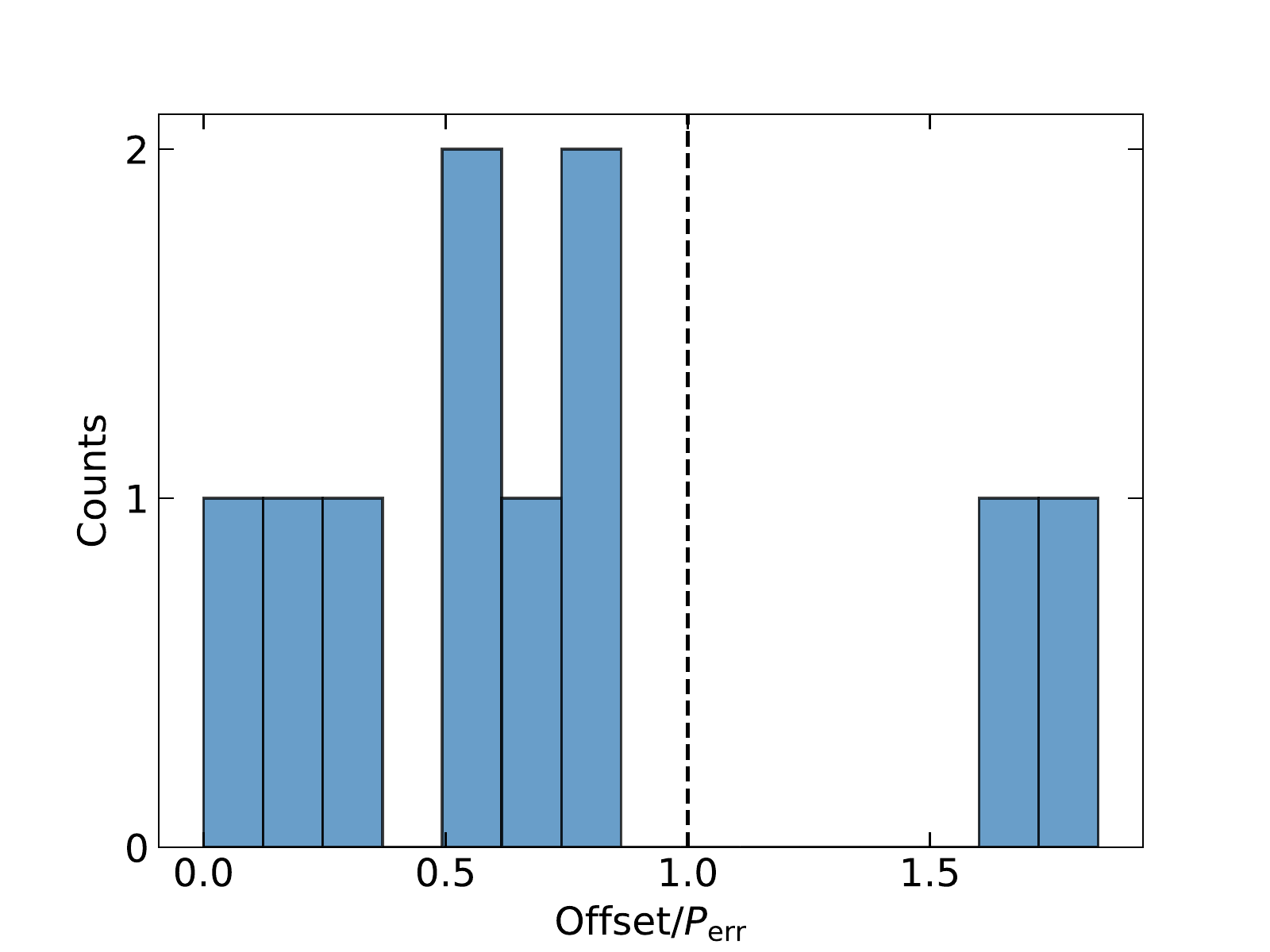}
     \caption{A histogram showing the distribution of offsets for the 10 identified couterparts. Each offset was normalised to the corresponding 95\% error radius ($P_\mathrm{err}$). The dashed line marks edges of the error circles.}
     \label{fig:offset_distribution}
 \end{figure}

\begin{figure*}
    \centering
    \includegraphics[scale=0.65]{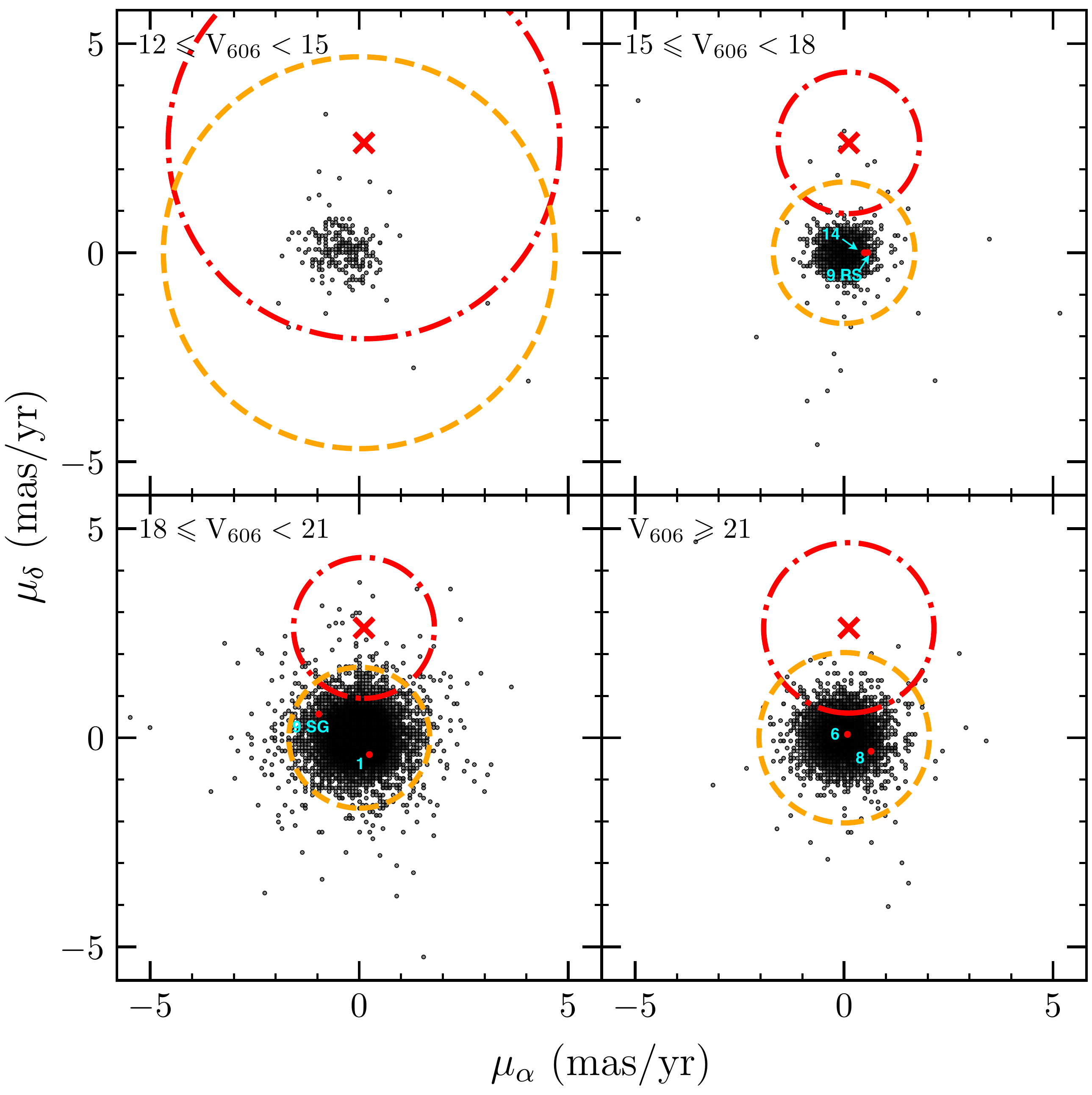}
    \caption{Proper motion distribution of stars with at least one good photometric measurement in each filter from the Padova Catalogue plotted in 4 $\mathrm{V_{606}}$ magnitude bins. The cluster's average proper motion corresponds to the zero point. The proper motion of background galaxies, obtained from GAIA DR2's estimate of the cluster motion, is marked with a red cross. Counterparts are shown with red points. The orange dashed circle in each panel shows the $3$ rms radius within which stars are considered as likely cluster members. The red dotted-dashed circle in each panel shows the corresponding $3$ $\sigma$ composite error radius of the AGN proper motion.}
    \label{fig:vpd}
  \end{figure*}

 Relying purely on photometric properties, we found potential counterparts to 10 of 16 {\it Chandra} sources. 8 of these are located within the corresponding 95\% error circles, all except for the possible counterparts to CX7 and CX16. Fig. \ref{fig:offset_distribution} shows a histogram of offsets for 10 identified counterparts in unit of their 95\% error radii. PMs for our proposed counterparts to CX2, CX7, CX12, CX13 and CX16 are not available due to their nondetections in 
 either 2006 or 2012, or both (see Tab. \ref{tab:hst_counterparts}). However, the counterparts to CX1, CX6, CX8, CX9 and CX14 all have PMs less than 3 rms different from M3 (both in the $\alpha$ and $\delta$ directions), and are inconsistent with the expected location of AGN, %on the PM plane: all reside within the 3 rms PM radius consistent with cluster membership, but 
 outside the expected $3$ $\sigma$ PM radius expected for background sources. The cluster memberships of the counterparts to CX1, CX6, CX9 and CX14 are confirmed by the calculated membership probability ($P_\mu$). The counterpart to CX8 was detected as a very faint extension to a relatively bright star in the 2006 ACS/WFC observations (see Fig. \ref{fig:finding_charts}), which might cause ambiguity in measuring the positions, so the PM information should be taken with care. %{\color{red} Perhaps we can include a strong statement here, e.g. at greater than 3 sigma confidence, depending on the uncertainty in the Gaia proper motions of the GC.}
 Fig. \ref{fig:finding_charts} shows a mosaic of finding charts, wherein the most likely counterpart to each {\it Chandra} source is annotated with a red arrow. The corresponding optical colours and magnitudes of these stars are marked with red circles in Fig. \ref{fig:fuvcmd}. The magnitudes used for identification are summarised in Tab. \ref{tab:hst_counterparts}.
 
 \subsection{Cluster Subpopulations and Chance Coincidences}
 \label{sec:chance_coincidence}
 In crowded regions (e.g. the cluster core), %some identified counterparts are more likely to be chance coincidences with cluster members,
 the chance of a potential counterpart being a coincidence is significant, 
 so it is important to estimate the number of chance coincidences for different cluster subpopulations. To get a census of cluster members, we used the PM information in the Padova catalogue to remove the  non-members. We found %a total of 
 $\approx 45554$ cluster members detected in the WFC3 FOV. We applied polygonal selection areas on the $\mathrm{UV-NUV}$ CMD to divide members into different subpopulations (see Fig. \ref{fig:subpopulations}). We found $\approx 2592$ evolved stars, including $\approx 967$ subgiants, and $\approx 1625$ red giants. %$\approx$ 
 108 stars were identified as (moderately) blue stars.  Finally, we found a population of 169 blue stragglers and a population of only 9 red stragglers.
 The predicted number of chance coincidences per error circle was %roughly
 estimated by multiplying the number of stars in each subpopulation by the ratio of the average area of the error circles ($\approx 0.72~\mathrm{arcsec^2}$) to the WFC3 FOV. %We found the predicted number of finding 
  We found that the average error circle contains $\approx 1.25$ normal cluster stars, though this varies somewhat by location in the cluster (see Fig. \ref{fig:finding_charts}). Among evolved members, %we found the number to be 
 an average error circle contains $\approx 0.04$ red giants and $\approx 0.03$ subgiants, so there is a substantial chance of finding a giant and/or subgiant star within an error circle. %We found only minuscule numbers for 
 However, the probability of finding a 
 blue star in an error circle is quite small ($\approx 3.01\times 10^{-3}$), as is the probability of a red straggler star ($\approx 2.47\times 10^{-4}$) or a blue straggler star ($\approx 4.63\times 10^{-3}$). Thus, the probability of finding a blue star in any of the 16 error circles by chance is only 5\%, and that of finding a red straggler is only 0.4\%. The probability of finding a blue straggler by chance coincidence is comparable to that of finding a blue star ($\approx 7\%$).
 
  \begin{figure*}
    \centering
    \includegraphics[scale=0.275]{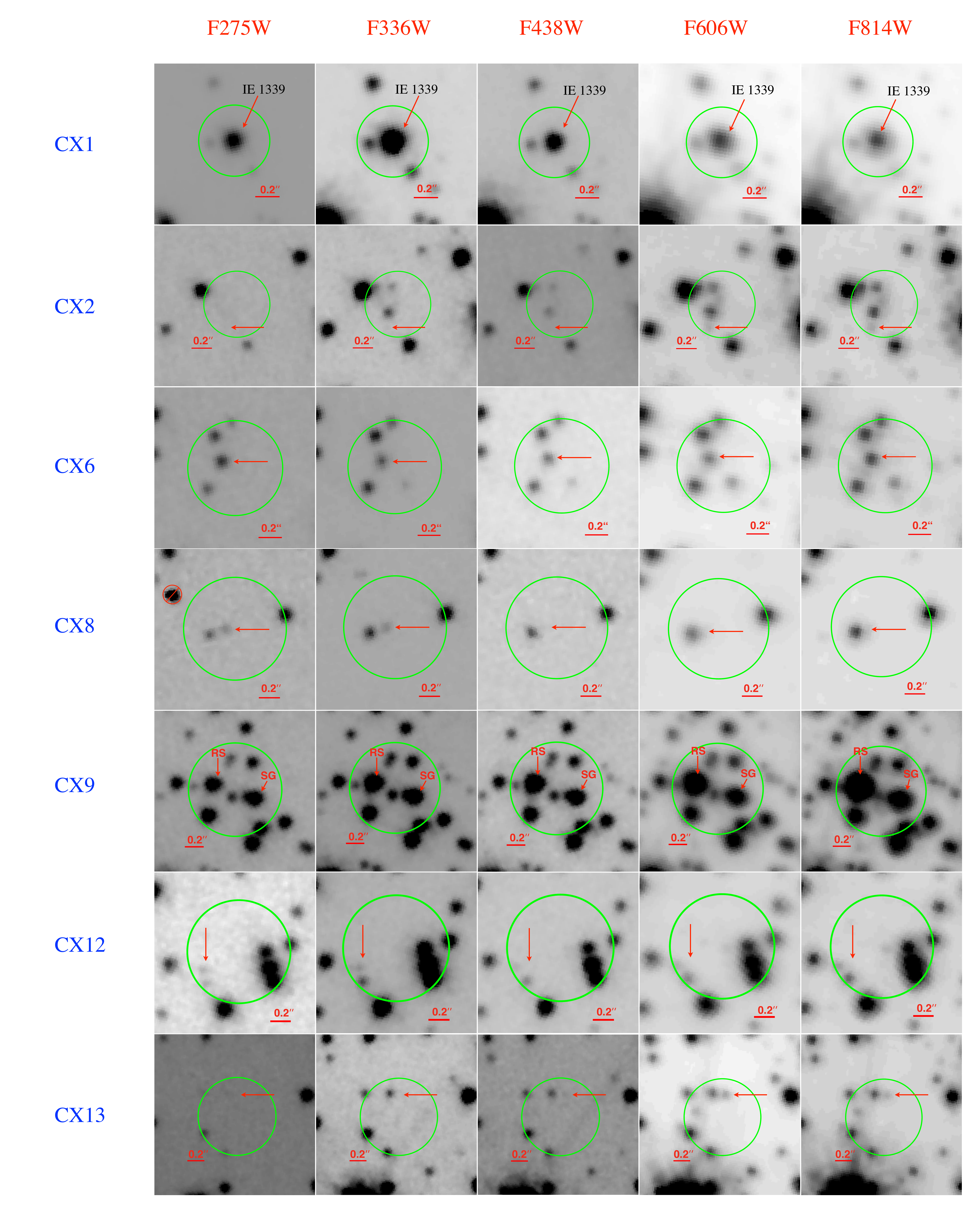}
    \caption{Finding charts for 7 identified optical counterparts in 5 different bands (UV bands are from the 2012 WFC3/UVIS observations; $\mathrm{V_{606}}$ and $\mathrm{I_{814}}$ bands are from the 2006 ACS/WFC observations) made from drizzle-combined images. North is up and east is to the left. The green solid circles represent the 95\% {\it Chandra} error circles as calculated in \citet{hong2005}. Identified counterparts are annotated with red arrows. Note that the counterparts to CX12 was only found in the two UV filters ($\mathrm{UV_{275}}$, $\mathrm{NUV_{336}}$); similarly, the counterparts to CX2 and CX13 were only detected in the two ACS/WFC bands, so the red arrows on the other finding charts only point to their nominal positions. 
    %This is similar for CX16, but only a less certain (outside the 90\% {\it Chandra} error circle) counterpart was found. 
    The counterpart to CX8 appears to be a faint extension in $\mathrm{V_{606}}$ and $\mathrm{I_{814}}$. Also, notice that a cosmic ray on the $\mathrm{UV_{275}}$ finding chart of CX8 has been excluded to avoid confusion. The red straggler (RS) and subgiant (SG) potential counterparts to CX9 are annotated with arrows and texts. }
    \label{fig:finding_charts}
  \end{figure*}

  \begin{figure*}
    \centering
    \includegraphics[scale=0.285]{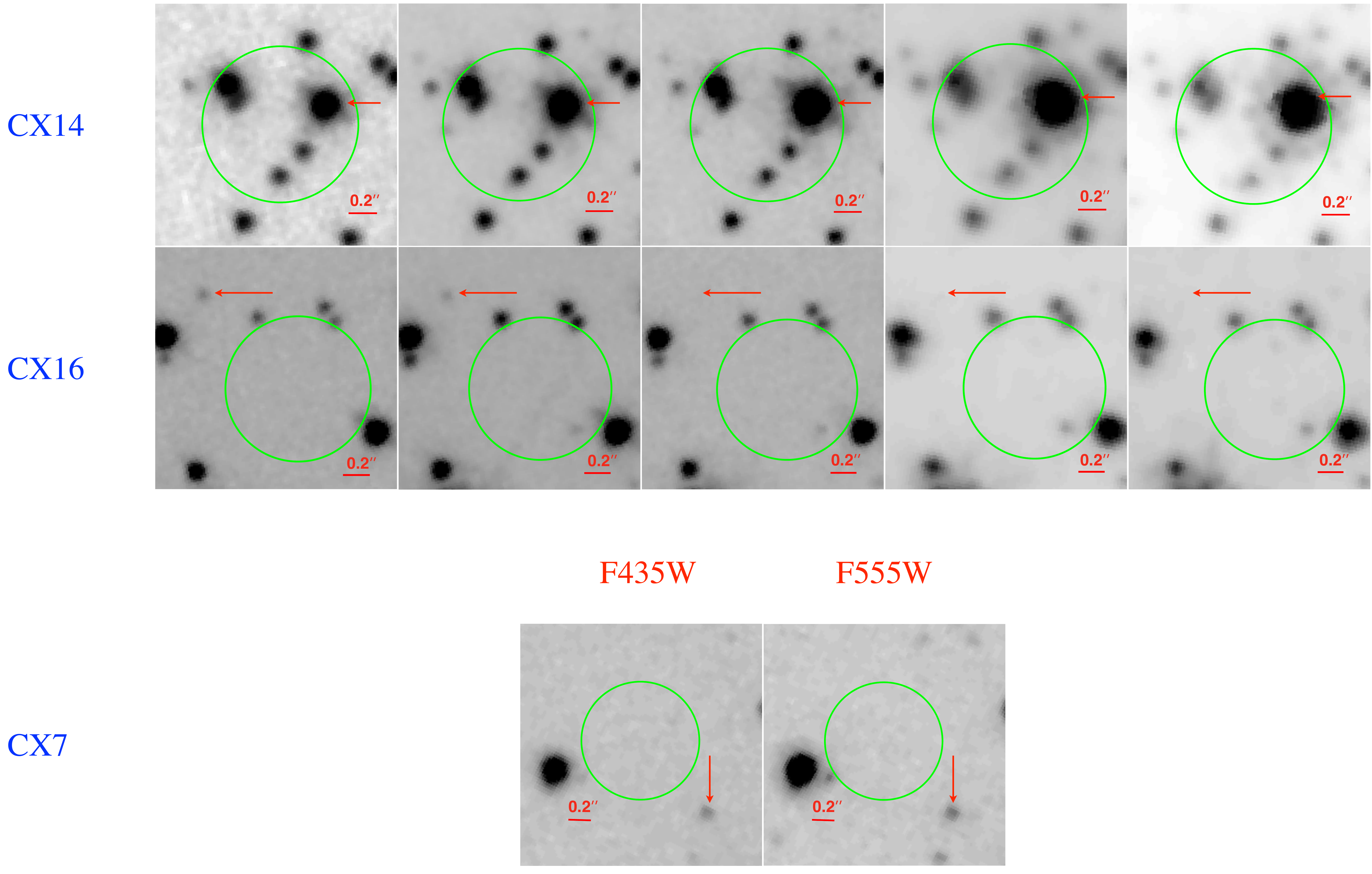}
    \contcaption{CX7 was only covered by the 2004 ACS/WFC observations. Plausible counterparts for CX7 and CX16 lie somewhat outside the 95\% {\it Chandra} error circle.}
    \label{fig:finding_charts_continued}
  \end{figure*}

  \begin{figure}
      \centering
      \includegraphics[scale=0.43]{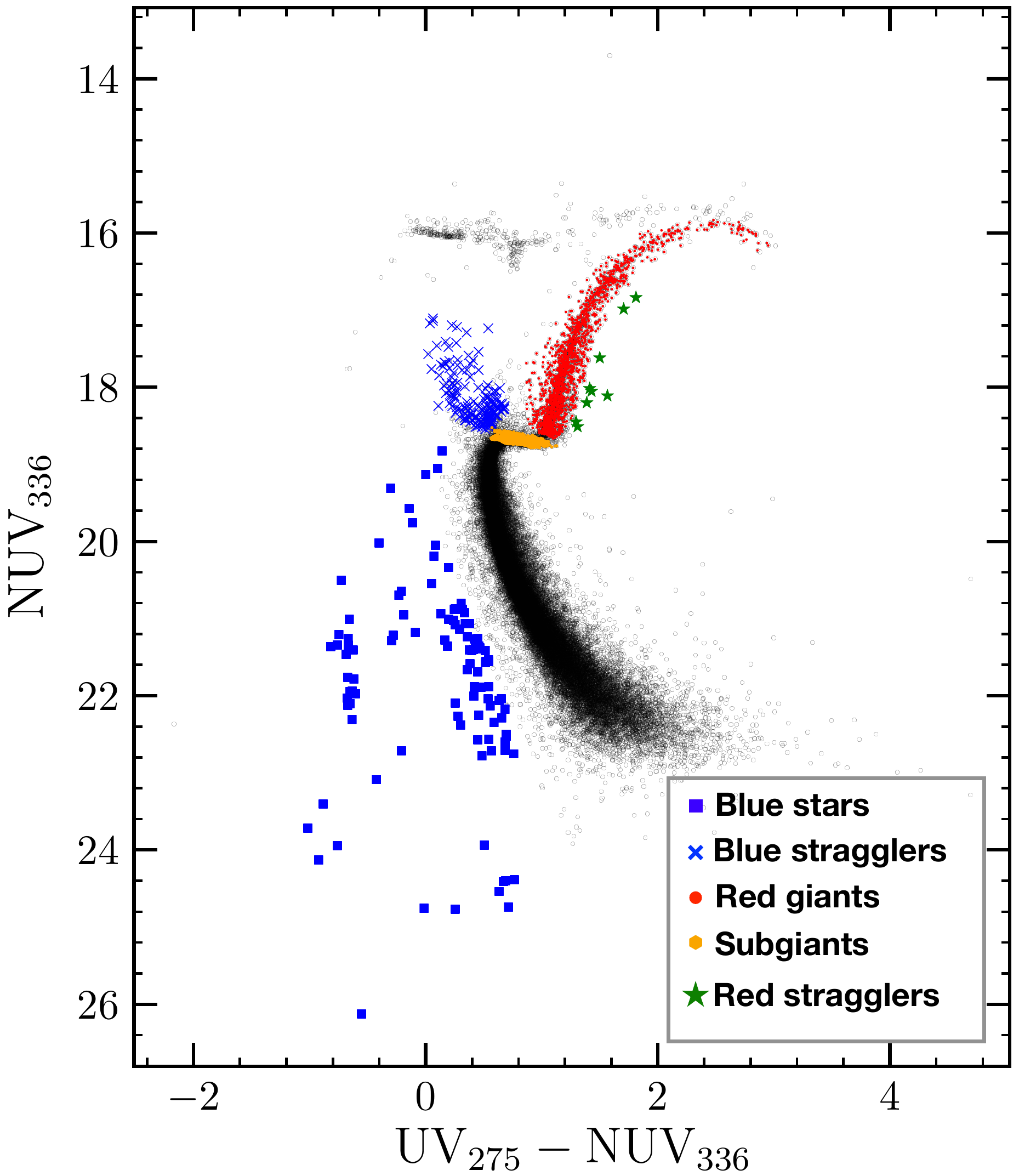}
      \caption{Proper motion cleaned $\mathrm{UV_\mathrm{275}-NUV_\mathrm{336}}$ CMD overplotted with different subpopulations.}
      \label{fig:subpopulations}
  \end{figure}

%{
%  \renewcommand{\arraystretch}{3}
  \begin{table*}
    \centering
    \caption{Magnitudes and colours of identified optical counterparts.}
    \resizebox{\textwidth}{!}{%
    \begin{tabular}{cccc|cc|cc|cc}
    \hline
    \hline
         CX  & \multicolumn{7}{c}{Magnitudes$^a$} &\multicolumn{2}{c}{Notes}\\
    \hline
       & \multicolumn{3}{c}{2012} & \multicolumn{2}{c}{2006} & \multicolumn{2}{c}{2004} &  \\
    \hline
       & $\mathrm{UV_{275}}$ & $\mathrm{NUV_{336}}$ & $\mathrm{B_{438}}$ & $\mathrm{V_{606}}$ & $\mathrm{I_{814}}$ & $\mathrm{B_{435}}$ & $\mathrm{V_{555}}$ & $P_\mu^b$ & Comments\\
    \hline
       1 & $18.02\pm 0.01$ & $17.88\pm 0.02$ & $18.81\pm 0.01$ & $18.05\pm 0.03$ & $17.37\pm 0.04$ & $18.14\pm 0.18$ & $18.10\pm 0.18$ & $96.9\%$ & \makecell[l]{Blue in UV,\\ moderate red excess in V-I}\\[0.4cm]
       2 & - & - & - & $23.66\pm 0.03$ & $22.50\pm 0.05$ & $25.49\pm 0.48$ & $24.08\pm 0.05$ & - & \makecell[l]{Faint MS star, \\ small red excess}\\[0.4cm]
       6 & $22.08\pm 0.01$ & $22.13\pm 0.02$ & $22.06\pm 0.02$ & $21.32\pm 0.02$ & $20.30\pm 0.05$ & $21.71\pm 0.22$ & $21.42\pm 0.00$ & $98.0\%$ & \makecell[l]{Blue in UV, \\ large red excess in V-I} \\ [0.4cm]
       7$^\dag$ & - & - & - & - & - & $25.09\pm 0.33$ & $24.78\pm 0.01$ & - & \makecell[l]{Faint blue star \\ outside the error circle} \\[0.4cm]
       8 & $23.34\pm 0.03$ & $23.14\pm 0.03$ & - &$24.37\pm 0.16$ & $23.63\pm 0.29$ & - & - & - & \makecell[l]{Faint blue star} \\ [0.4cm]
       9 & $19.91\pm 0.02$ & $18.48\pm 0.02$ & $18.30\pm 0.01$ & $16.88\pm 0.00$ & $16.14\pm 0.00$ & $18.09^\ast$ & $17.03\pm 0.06$ & $88.9\%$ & \makecell[l]{Red straggler} \\ [0.4cm]
       9 & $19.78\pm 0.02$ & $18.96 \pm 0.02$ & $19.12 \pm 0.02$ & $18.31 \pm 0.01$ & $17.77\pm 0.01$ & $18.98\pm 0.00$ & $18.51\pm 0.00$ & $97.7\%$ & \makecell[l]{Subgiant} \\[0.4cm]
       12 & $24.37\pm 0.03$ & $24.30\pm 0.04$ & - & - & - &- &- & - & \makecell[l]{Faint blue star, \\only detected in UV} \\ [0.4cm]
       13 &- &- &- & $22.72\pm 0.08$ & $22.64\pm 0.29$ & $23.90^\ast$ & $23.37\pm 0.72$ & - & \makecell[l]{Moderately blue star}\\[0.4cm]
       14 & $18.63\pm 0.02$ & $16.97\pm 0.01$ & $16.76\pm 0.01$ & $15.51\pm 0.013$ & $14.79\pm 0.04$ & - & -  & $96.6\%$ & \makecell[l]{Red giant} \\[0.4cm]
       16$^\dag$ & $23.75\pm 0.02$ & $24.15\pm 0.02$ & - & - & - & $25.69\pm 0.32$ & $26.34\pm 0.89$ & - & \makecell[l]{Faint blue star \\ outside the error circle,\\ only detected in UV}  \\
    \hline
    \multicolumn{6}{l}{$^a$ Magnitudes are calibrated to VEGAMAG system}\\
    \multicolumn{6}{l}{$^b$ Probability of being a member of the cluster, from \citet{nardiello2018}}\\
    \multicolumn{6}{l}{$^\ast$ superscript indicates that the magnitude has a very large uncertainty and should be taken with care}\\
    \multicolumn{6}{l}{$^\dag$ superscript indicates that the optical counterpart is outside the 95\% {it Chandra} error circle}
    \end{tabular}}
    \label{tab:hst_counterparts}
\end{table*}
%}

\subsection{Optical Variability}
 Signs of optical/UV variability can be helpful to further confirm the nature of the source, especially for CV identifications, since most CVs appear as blue variable stars (see e.g. \citealt{Cool98,Edmonds03b,Dieball17,RiveraSandoval18}). 

\begin{figure*}
    \centering
    \includegraphics[scale=0.34]{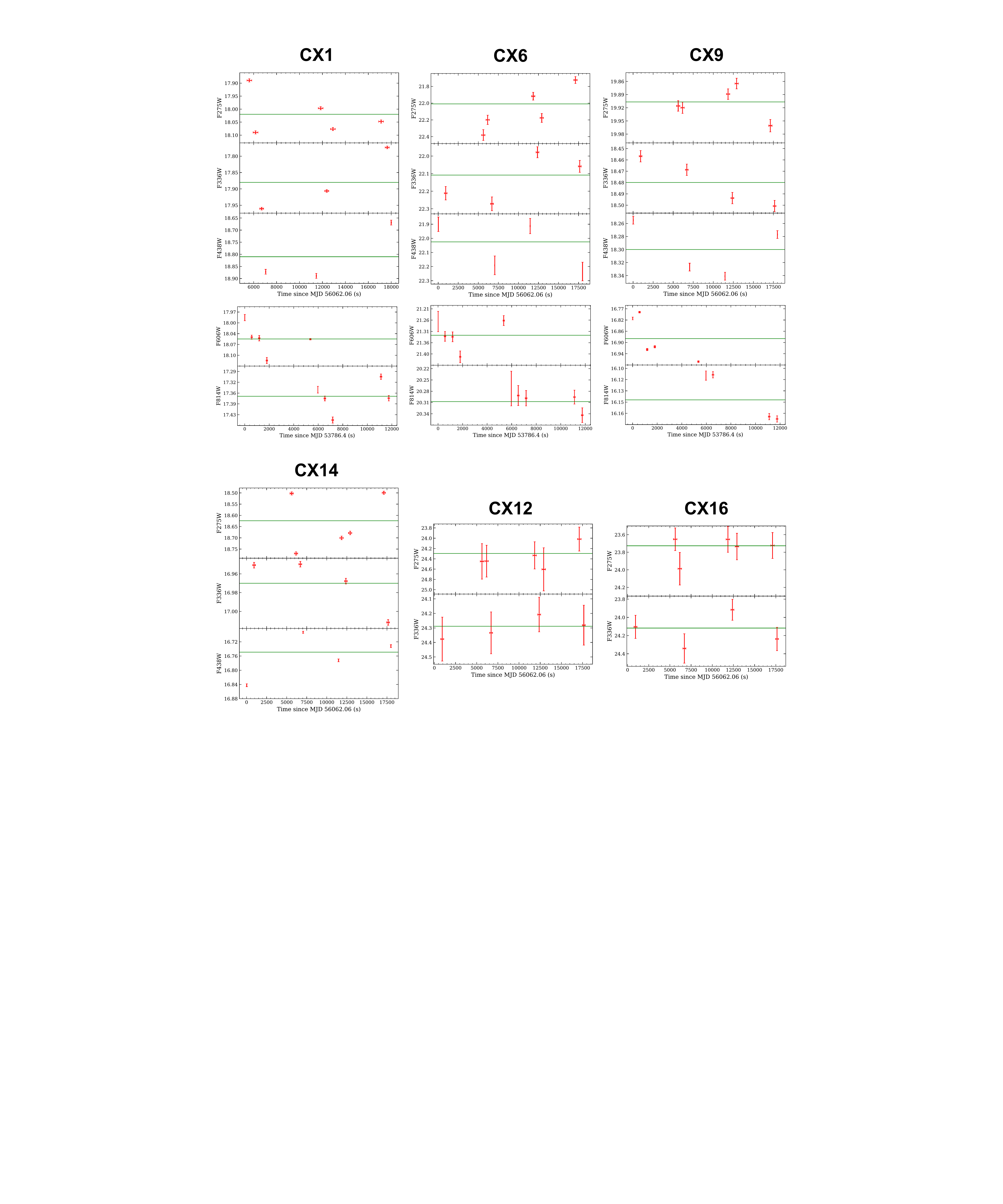}
    \caption{Optical/UV light curves of the identified counterparts that have detections in multiple exposures in the WFC3/UVIS observations (2012) and/or the ACS/WFC observations (2006). Photometric errors are from DAOPHOT software. The best fit constants are indicated with a solid green line.}
    \label{fig:light_curve_uv}
\end{figure*}

 For the new counterparts that were detected in multiple exposures, we constructed {\it HST} light curves in the 2012 WFC3/UVIS and the 2006 ACS/WFC bands by doing DAOPHOT photometry (aperture photometry if the surrounding field of the counterpart is relatively uncrowded, e.g. CX16. Otherwise, PSF-fitting photometry was applied, e.g. CX9) on individual CR-removed, distortion-corrected FLC frames. The resulting light curves are shown in Fig. \ref{fig:light_curve_uv}. We then used least-square fitting to fit each light curve to a constant and used the resulting reduced $\chi^2$ (on Fig. \ref{fig:light_curve_uv}) as a measure of variability. The resulting $\chi^2$ are summarised in Tab. \ref{tab:variability}. 
 
 \begin{table}
     \centering
     \begin{tabular}{c|ccc|cc}
     \hline
     \hline
        CX  & \multicolumn{5}{c}{Filters} \\
            & \multicolumn{3}{c}{2012} & \multicolumn{2}{c}{2006}\\
     \hline
            & $\mathrm{UV_{275}}$ & $\mathrm{NUV_{336}}$ & $\mathrm{B_{435}}$ & $\mathrm{V_{606}}$ & $\mathrm{I_{814}}$ \\
     \hline
         1  & $183.1$ & $578.9$ & $149.8$ & $24.4$ & $29.6$ \\
         6  & $27.0$  & $14.9$  & $8.9$   & $6.1$ & $1.2$ \\
         9  & $7.7$   & $17.2$  & $46.0$  & $544.3$ & $53.3$ \\
         12 & $0.6$   & $0.3$  & - & - & -\\
         14 & $296.9$ & $93.7$  & $313.4$ & - & - \\
         16 & $0.6$   & $2.0$ & - & - & -\\
     \hline
     \end{tabular}
     \caption{Summary on resulting $\chi^2$ from fitting counstants to light curves in different filters.}
     \label{tab:variability}
 \end{table}
 
 %The X-ray variability was analysed by using the CIAO {\it glvary} tool on each of the three observations. The {\it glvary} applies the Gregory-Loredo algorithm \citep{greg1992} that assigns variability indices and based on differences between time bins. Besides the clear variability in CX1 \citep[variability indices of 7 and 6 in the first and second observation, respectively]{stacey2011}, CX5 and CX6 show strong sign of variability in the first and second {\it Chandra} observation, respectively (corresponding to variability index of 5 and 6). 

\subsection{SED-Modelling of 1E1339}
\label{sec:sed_fitting}
%{\color{red} Suggest moving to section 4.1.  Suggest adding a new, small table for these magnitudes.}
 As mentioned above, the 2004 ACS data for 1E1339 (see Tab. \ref{tab:hst_obs}) were taken over a broad range of filters, which yielded a dataset that can be used to construct and analyse the spectral energy distribution (SED). The target is well resolved and isolated in the three HRC filters, so that we can obtain photometric results with relatively high accuracy. However, the source is affected by saturation blemish from a nearby bright star on the $\mathrm{B_{435}}$, $\mathrm{V_{555}}$ and $\mathrm{I_{814}}$ images. We, therefore, performed photometry with relatively small apertures (radius of $0.05\arcsec$ or 1 ACS/WFC pixel) for these three filters. To calculate fluxes, we first calculated count rates from instrumental magnitudes. These count rates were then calibrated to infinite apertures by dividing the corresponding encircled energy (EE) fractions from \cite{sirianni2005}.
 Finally, count rates were converted to specific fluxes ($F_\lambda$) by multiplying the corresponding inverse sensitivities (PHOTFLAM\footnote{\url{http://www.stsci.edu/hst/acs/analysis/}}), i.e.

 \begin{equation}
     F_\lambda = \dfrac{\texttt{Count rate in a small aperture}}{\texttt{EE}}\times \texttt{PHOTFLAM}.
     \label{eq1}
 \end{equation}

 The specific fluxes in different filters are summarised in Tab. \ref{tab:cx1_fluxes}. The calibrated $\mathrm{B_{435}}$ and $\mathrm{V_{555}} $magnitudes are also summarised in Tab. \ref{tab:hst_counterparts}.
 
 \begin{table*}
     \centering
     \caption{Calibrated specific fluxes ($F_\lambda$) of CX1 in different filters from the 2004 ACS observation.}
     \begin{tabular}{ccccccc}
     \hline
      & \multicolumn{3}{c}{ACS/HRC} & \multicolumn{3}{c}{ACS/WFC} \\
      Filters & F220W  & F250W  & F330W & F435W & F555W  & F814W  \\
      \hline
        Flux ($\times 10^{-16}~\mathrm{erg~s^{-1}~cm^{-2}~\angstrom^{-1}}$) & $6.62(43)$ & $7.92(53)$ & $7.36(50)$ & $4.07(30)$ & $2.83(14)$ & $2.02(10)$\\
    \hline
     \end{tabular}
     \label{tab:cx1_fluxes}
 \end{table*}

\begin{figure*}
    \centering
    \includegraphics[scale=0.6]{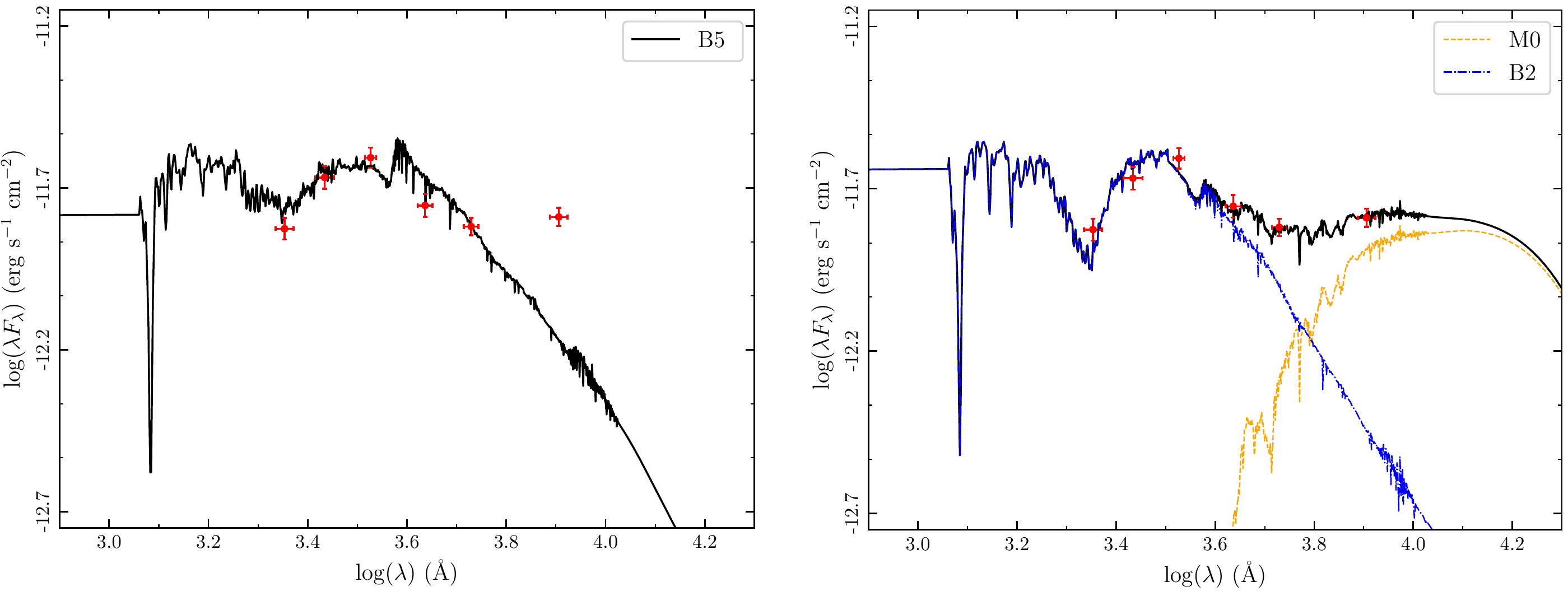}
    \caption{Best fit SED models and data from the 2004 observations. {\it Left panel} shows the best fit single-component model (solid black) overplotted with the data (red). {\it Right panel} shows the same data (red) and the best fit two-component model (solid black), composed of a renormalised B2 component (blue dashed-dotted) and an renormalised M0 component (orange dashed). }
    \label{fig:cx1_sed}
\end{figure*}

 We used tabulated stellar SED models from the Pickles library, an atlas of 131 stellar spectral models, \citep{pickles1998} to model the SED. Spectra were convolved with the filter bandpasses and corrected for the expected extinction using PySynphot package\footnote{\url{http://pysynphot.readthedocs.io/en/latest/}}. The model spectra were then renormalised with a $\chi^2$ minimization process (see Appendix). We first tried $\chi^2$ fits with one-component models. The minimum of $\chi^2$ was obtained with a type B5I spectrum, but still a bad fit ($\chi^2\approx 153.89$). This was mainly caused by an obvious excess in the data (vs. the models) in the $\mathrm{I_{814}}$ band ($\log(F_{\lambda, \mathrm{data}}) - \log(F_{\lambda, \mathrm{model}}) \approx 0.37$, corresponding to a difference in magnitude of $\approx 0.92$), requiring a second cooler component to compensate for the excess. We, therefore, tried with two-component composite models. This is done by looping over all possible combinations of two components drawn from the Pickles library and picking out the combination with the smallest $\chi^2$. The best fit was obtained with a B2 + M0 model, which is a greatly improved fit ($\chi^2\approx 3.42$). The M0 component now accounts for the V and I excesses relative to the B2 component. %With one component fixed, we tried to constrain the confidence interval of the spectral types and therefore the effective temperatures. 
 To constrain the spectral types of each component, we stepped one component at a time through different spectral types, allowing the other fit parameters to vary.
 We found a 90\% confidence interval of K3-M2 or $T_\mathrm{eff} = 3.75_{-0.15}^{+1.05}\times 10^3~\mathrm{K}$ for the redder component. For the bluer component, we %obtained a less constrained confidence interval (>90\%) of 
 found 
 O8-B5 or $T_\mathrm{eff} = 2.10^{+1.96}_{-0.58}\times 10^4~\mathrm{K}$.
 The data overplotted with the best fit models are shown in Fig. \ref{fig:cx1_sed}.
 
 Care should be taken, however, that the above spectral types do not necessarily reflect the actual traits of the accretor or the donor. Typically in CVs, the O or B type spectral components are ascribed to the combined light from the accretion disc and/or the WD surface, while the M type component might be caused by binary interactions. We will further discuss this in Sec. \ref{sec:sec_cx1}
 
%which are available at the ACS Data Analysis webpage\footnote{\url{http://www.stsci.edu/hst/acs/analysis/}}. Finally, count rates were multiplied by the corresponding inverse sensitivities.

%Finally, count rates were multiplied by the corresponding inverse sensitivities.

%Finally, the count rates were multiplied by the inverse sensitivities , converting them to specific fluxes

%  and were finally converted into specific fluxes ($F_\lambda$) using the inverse sensitivities (PHOTFLAM) as conversion factors, which are available at the ACS Data Analysis webpage (PHOTFLAM) . Figure shows the resulting SED.

\subsection{X-ray spectral analysis}
 We performed detailed X-ray spectral analysis for sources with $>$100 counts that have not previously been published; only one source in our {\it Chandra} catalogue fits this description, CX2. 
 For other sources with $>$10 counts, we perform simplified spectral analyses. We used the CIAO {\it specextract}\footnote{\url{http://cxc.cfa.harvard.edu/ciao/threads/pointlike/}} script to extract spectra of these sources from each of the 3 observations. Because these observations do not span a very long time, to get better statistics, we combined the 3 spectra and their responses using FTOOLS/ADDSPEC\footnote{\url{https://heasarc.gsfc.nasa.gov/ftools/caldb/help/addspec.txt}} before doing the analysis. The combined spectra were then re-binned using the CIAO {\it dmgroup}\footnote{\url{http://cxc.harvard.edu/ciao/ahelp/dmgroup.html}} tool to at least 1 count per bin. Spectral analysis was then performed on the combined spectra with HEASoft/Xspec version 12.9.1 using C-statistics \citep{cash1979}. Because the accuracy of the {\it Chandra} response matrix file falls off at low energies, %due to the low sensitivity of ACIS-S detectors,
 we ignored energy channels below $0.5~\mathrm{keV}$ during the fits. (These channels are included in the plots just to demonstrate the obvious excess at low energies.) We fixed the absorption hydrogen column density at the cluster value $\sim 8.7\times 10^{19}~\mathrm{cm^{-2}}$, which is derived by applying the conversion factor ($\sim 2.81\pm 0.13 \times 10^{21}$) from \citet{Bahramian15} and using $E(B-V)=0.01$ from \citet[2010 edition]{harris1996}.

 %Since CX2 has relatively more counts, 
 We tried fits to CX2 with multiple models including an empirical power law (POWERLAW), a blackbody (BBODYRAD), a neutron star atmosphere (NSATMOS), and a hot plasma model (VMEKAL). For sources with more than $10$ counts but fainter than CX2, we fit an individual power law to each source. We combined the spectra of sources with less than $10$ counts (CX7-CX16) and fit a power law to the combined spectrum to get an average photon index. We obtained an average index of $\Gamma \approx 1.3$, which was then applied to the spectra of CX7-CX16 to calculate fluxes (fluxes of sources fainter than CX9 were calculated with {\it srcflux}, so no C-statistic or goodness calculation is available). All fitting results are summarised in Tab. \ref{tab:fitting_results}.

\begin{table*}
    \centering
    \caption{Summary of X-ray spectral analyses.}
    \resizebox{\textwidth}{!}{%
    \begin{tabular}{ccccccccc}
    \hline
    \hline
    Source & Model & $n_H^a$ & $\Gamma$ or $R_i^b$ & $kT$ & $f_{0.5-7.0, \mathrm{unabs}}$ & $L_{0.5-7.0, \mathrm{unabs}}$ & Cstat/dof & Goodness\\
    CX     & & $(10^{19}~\mathrm{cm^{-2}})$  & & $\mathrm{(keV)}$ & $\mathrm{(10^{-16}~erg~cm^{-2}~s^{-1})}$& $\mathrm{(10^{30}~erg~s^{-1})}$ & &\\
    \hline
    \multirow{4}{*}{2} & TBabs*POWERLAW & \multirow{4}{*}{$(8.7)$} & $3.6^{+0.4}_{-0.3}$ & - & $206.6^{+30.1}_{-27.4}$ & $257.2^{+37.5}_{-34.2}$ & $53.1/59$ & $66.8\%$ \\ 
                       & TBabs*BBODYRAD &  & $1.4^{+0.5}_{-0.3}$ & $0.20^{+0.02}_{-0.02}$  & $177.4^{+25.9}_{-23.6}$ &      $220.9^{+32.2}_{-29.4}$ & $49.3/59$ & $53.8\%$ \\
                       & TBabs*VMEKAL   &  & - & $0.42^{+0.11}_{-0.06}$ & $184.1^{+26.8}_{-24.5}$ & $229.2^{+33.4}_{-30.5}$ & $60.4/59$ & $92.9\%$ \\
                       & TBabs*NSATMOS  &  & $8.6^{+5.7}_{-3.6^\ast}$ & $0.10^{+0.06}_{-0.02}$ & $183.3^{+26.7}_{-24.4}$ & $228.2^{+33.3}_{-30.3}$ & $46.2/59$ & $9.6\%$\\
    \hline
    3                  & \multirow{4}{*}{TBabs*POWERLAW} & \multirow{4}{*}{$(8.7)$} & $0.8^{+0.4}_{-0.4}$ & \multirow{4}{*}{-} & $167.5^{+45.0}_{-38.1}$ & $208.6^{+56.1}_{-47.5}$ & $43.1/34$ & $51.9\%$ \\
    4                  & & & $0.8^{+0.6}_{-0.6}$ & & $63.9^{+27.3}_{-21.4}$  & $79.5^{+34.0}_{-26.6}$ & $11.08/15$ & $3.3\%$ \\
    5                  & & & $2.1^{+0.8}_{-0.7}$ & & $43.3^{+17.6}_{-13.8}$ & $53.9^{+21.9}_{-17.2}$ & $21.6/17$ & $29.2\%$\\
    \multirow{2}{*}{6}              & & & $-0.7^{+0.7}_{-0.8}$& & $104.9^{+50.2}_{-38.1}$ & $130.6^{+62.4}_{-47.5}$ & $18.1/13$ & $6.7\%$\\
                       & & $(6\times 10^3)$ & $1.2^{+1.2}_{-1.3}$& & $191.2^{+93.0}_{-70.4}$ & $238.1^{+115.7}_{-87.7}$ & $20.1/13$ & $25.8\%$\\
    \hline
    7                    & \multirow{10}{*}{TBABS*POWERLAW} & \multirow{10}{*}{$(8.7)$} & \multirow{10}{*}{$1.3^{+0.4}_{-0.4}$} & \multirow{10}{*}{-} & $41.7^{+22.2}_{-16.5}$ & $52.0^{+27.7}_{-20.5}$ & $10.7/15$ & $66.9\%$\\
    8                    &  &  &  &  & $27.5^{+19.5}_{-13.3}$ & $34.2^{+24.3}_{-16.6}$ & $3.4/7$ & $33.1\%$ \\
    9                    &  &  &  &  & $20.0^{+16.8}_{-11.0}$ & $24.9^{+21.0}_{-13.7}$ & $1.0/6$ & $8.5\%$ \\
    10                   &  &  &  &  & $18.9^{+17.5}_{-8.6}$ & $23.5^{+21.7}_{-10.7}$ &- &-\\
    11                   &  &  &  &  & $17.0^{+17.3}_{-8.1}$ & $21.1^{+21.5}_{-10.1}$ & - &-\\
    12                   &  &  &  &  & $14.8^{+16.9}_{-7.6}$ & $18.4^{+21.0}_{-9.5}$ & -&-\\
    13                   &  &  &  &  & $15.9^{+16.8}_{-7.5}$ & $19.8^{+20.9}_{-9.4}$ & -&-\\
    14                   &  &  &  &  & $13.3^{+16.2}_{-6.7}$ & $16.5^{+20.2}_{-8.4}$ & -&-\\
    15                   &  &  &  &  & $11.0^{+12.1}_{-6.9}$ & $13.6^{+15.0}_{-8.6}$ & -&-\\
    16                   &  &  &  &  & $10.5^{+13.5}_{-6.2}$ & $13.0^{+16.7}_{-7.7}$ & -&-\\
    \hline
    \multicolumn{7}{l}{$^a$ Values in the parentheses indicates that the parameter is fixed during the fit}\\
    \multicolumn{7}{l}{$^b$ $R_i$ is the radius of the emission region for BBODYRAD model, or the neutron star radius for the NSATMOS model, both in km}\\
    %\multicolumn{7}{l}{$^c$ Unabsorbed fluxes were calculated by convolving models with $cflux$}\\
    \multicolumn{7}{l}{$^\ast$ indicates that the error extends beyond the hard limit}
    \end{tabular}}
    \label{tab:fitting_results}
\end{table*}

\section{Individual Sources \& Discussions}
\label{sec4}
\subsection{CX1/1E1339}
\label{sec:sec_cx1}
 %This source was first discovered as a faint X-ray source ($L_X \sim 10^{33}$) by the {\it Einstein Observatory} \citep{hertz1983}. It underwent a bright outburst when observed with ROSAT in 1991 and 1992, during which it showed a very soft spectrum ($kT \approx 20~\mathrm{eV}$, $L_X \sim 10^{35}$, \citealt{hertz1993}). The source returned back to quiescence with a much harder spectrum when observed with ASCA \citep{Dotani99} and {\it Chandra} ($\Gamma \sim 1.4$; see \citealt{stacey2011}). The bright outburst and the supersoft spectrum suggest a physical connection to other galactic supersoft X-ray sources (SSSs). However, 1E1339's peak observed X-ray luminosity was $\sim 100$ times fainter than that of standard SSSs, suggesting a much smaller burning area.  1E1339 is the only SSS identified in a GC so far.

 The optical counterpart of 1E1339 was first identified by \cite{edmonds2004}, using WFPC2 data, to be a star with a very blue $\mathrm{U-V}$ colour (but lying on the subgiant branch in a $\mathrm{V-I}$ CMD), showing marked variability on timescales of hours. Our photometry confirms its blue $\mathrm{U-V}$ excess. Though 1E1339 lies on the blue straggler region in  UV-NUV and $B-V$ CMDs, it does not lie in the blue straggler region in other CMDs (see Figures \ref{fig:fuvcmd}, \ref{fig:ksync_cmd}). This illustrates that multiple emission components are required, as expected in cataclysmic variables; examples of real blue straggler counterparts, and another CV that only appears to lie on the blue straggler sequence in some CMDs, can be found in \citet{Edmonds03a}.
 %this star is indeed very likely to be the bona-fide counterpart to CX1 according to our calculation of chance coincidences. 
 The ACS photometry from the {\it ACS Globular Cluster Treasury Program} revealed a red excess in $\mathrm{V-I}$ colour ($E(V-I) \approx 0.09$ with respect to the red giant branch), suggesting a red straggler (RS) secondary \citep{Mathieu03,Geller17}. %Simultaneous blue and red excesses 
 Unusually red secondaries
 in XRBs have been observed in some CVs in 47 Tuc \citep{Edmonds03a}, M30 \citep{lugger2007}, and NGC 6752 \citep{Thomson12}. 
 Such bloating may be due to irradiation, or more likely due to continued mass loss from the donor, as expected under standard CV evolutionary theory \citep{Knigge11}.
 %One possible scenario for this is that the envelope of the companion is irradiated by radiation from the compact object (a WD in this case) and/or the accretion disc such that it is extended and bloated, which therefore appears to be redder. 
 Using the correspondences between $T_\mathrm{eff}$ and $\mathrm{V-I}$ colours from \citet{cox2000}, we found that the envelope of 1E1339's companion should have expanded by roughly $3\%$ in order to have the observed red excess. \citet{leiner2017} and \citet{ivanova2017} show evolutionary models for binary mass-transfering systems that indeed demonstrate optical colours like these. This star also varies by $\approx 0.2~ \mathrm{mag}$s in the UV and $\mathrm{B}$ filters on timescales of hours.  Fitting its lightcurves to a constant gives a poor reduced $\chi^2$ (e.g. reduced $\chi^2\approx 183$ for the $\mathrm{UV_{275}}$ filter, see Fig. \ref{fig:light_curve_uv}).

%Although the data we used for SED modeling is a year apart from the 2006 ACS/WFC observations, the
 Our SED fitting result corroborates this observed red excess ($\approx 0.92~\mathrm{mags}$ as mentioned in Sec. \ref{sec:sed_fitting}), since a second, cooler component is required to get a satisfactory fit. The red excess might be partly due to variability, either long term or short term, of CX1.
 However, the presence of this red excess in two independent {\it HST} epochs (2004 and 2006 $V-I$ CMDs) indicates its extreme colour is likely real. The 0.92-mag difference between the observed and expected $I$ magnitudes is larger than can reasonably be explained by the observed short-term variability, of typically 0.2 magnitudes \citep{edmonds2004}.
 
 %We tested whether the longer-term variation observed by \citet{edmonds2004}, might explain this, adopting their linear fit in which $\mathrm{I_{814}}$ varies by $0.20\pm 0.04$ mags/day.
 %The 0.92-mag difference in this case is therefore hardly to be achieved solely by long term variability.
 %Over short time scales, the $\mathrm{I_{814}}$ magnitudes varies by $\approx 0.2$ mags (see Fig.\ref{fig:light_curve_uv}), which is also not strong enough to account for the observed red excess.
 %Therefore, the red excess is very likely to be a bona-fide second component.} 
 The cooler spectral component suggests an M0-type subgiant, while the bluer component has a spectral type for a B2-type giant, which is likely to represent the combined light from the WD and the accretion disc. We note that the extreme UV brightness indicates a high rate of mass transfer through the accretion disc. Should the accretor be a neutron star or black hole, a very high X-ray luminosity ($>10^{36}$ erg/s) would be expected; the lack of such bright X-rays indicates that the accretor is almost certainly a white dwarf.

\subsection{CX2--- a quiescent low-mass X-ray binary (qLMXB)?}
\label{sec:cx2}
 %Utilizing the relatively high X-ray count rate, we can constrain its spectral properties. 

\begin{figure}
    \centering
    \includegraphics[scale=0.5]{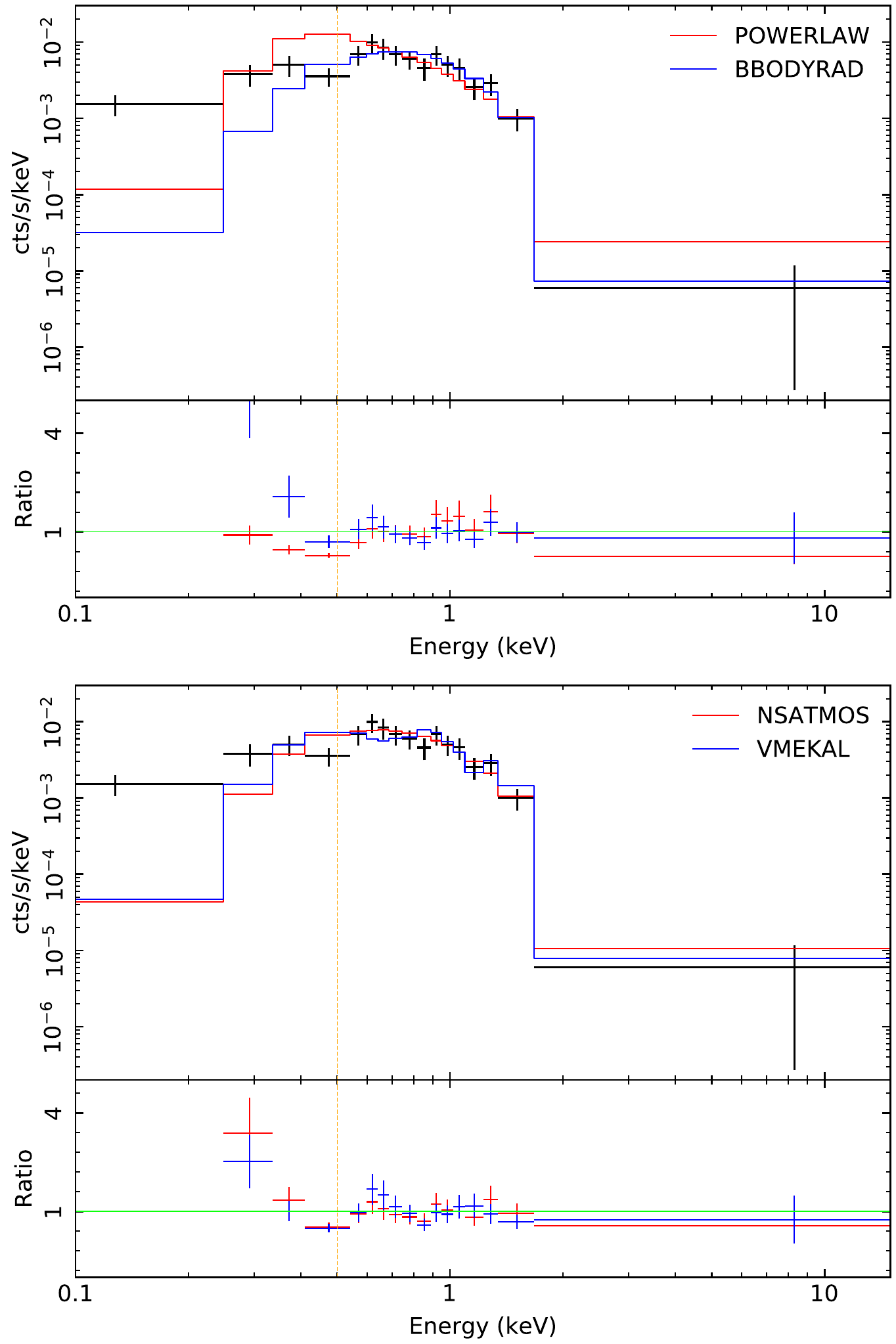}
    \caption{{\it Chandra} spectra of CX2. The top panel shows the rebinned (only for plotting purpose) data (black) with the best fit POWERLAW  model and BBODYRAD model overplotted with a solid red line and a solid blue line, respectively. The bottom panel shows the same data, but overplotted with the best fit VMEKAL model (blue) and NSATMOS model (red). The yellow dashed line indicates the energy limit at $0.5~\mathrm{keV}$, below which channels were ignored during the fits. %A low-energy excess is visible in all four models. 
    Ratio = data/model.}
    \label{fig:cx2_spec}
\end{figure}

 To model CX2's X-ray spectrum, we first tried a simple absorbed power-law model (TBabs*POWERLAW) and found a rather soft photon index ($\Gamma = 3.6^{+0.4}_{-0.3}$). The soft nature of this source was further confirmed by the low $kT$ ($\approx 0.2~\mathrm{keV}$) from a blackbody fit (TBabs*BBODYRAD). 
 We then tried fits with more physically motivated models. 
 A thermal plasma fit (TBabs*VMEKAL) yielded a slightly worse fit ($\mathrm{Goodness=92.9\%}$) with a $kT \sim 0.4~\mathrm{keV}$. 
  A neutron star atmosphere model  (TBabs*NSATMOS, \citealt{heinke2006}) consistently yielded a low $kT_\mathrm{eff}$ $\approx 0.10~\mathrm{keV}$ (typical for a quiescent low-mass X-ray binary (qLMXB)) with an NS radius $R_\mathrm{NS}\approx 8.6~\mathrm{km}$ ($\mathrm{Goodness=9.6\%}$), or $kT_\mathrm{eff}\approx 0.09$  when $R_\mathrm{NS}$ was frozen to $10~\mathrm{km}$ ($\mathrm{Goodness=13.2\%}$). The models and best fit parameters are summarised in Tab. \ref{tab:fitting_results}.
 Fig. \ref{fig:cx2_spec} shows the spectrum overplotted with the best fit models (see Tab. \ref{tab:fitting_results}). 
 %{\color{red} Should refer to the table of spectral fits here as well.} 
 Below 0.5 keV, the data are poor fits, but we attribute this to the difficulty of calibrating this portion of the X-ray spectrum. 

 We identify a potential optical counterpart to this source, a star that lies on the red side of the main sequence in the $\mathrm{B_{435}-V_{555}}$ CMD (Fig.~\ref{fig:ksync_cmd}). This star shows a relatively large photometric error in the $\mathrm{B_{435}}$ band (see Tab. \ref{tab:hst_counterparts}), which makes its CMD position uncertain. %With the population synthesis information provided in 
 However, as this counterpart is not clearly off the main sequence, Sec. \ref{subsec:counterpart_search} suggests that this star may well be a chance coincidence.
 % we estimated the number of chance coincidences within the {\it Chandra} error circle to be $\approx 0.56$, so the suggested counterpart might also be a chance coincidence.

\begin{figure}
    \centering
    \includegraphics[scale=0.36]{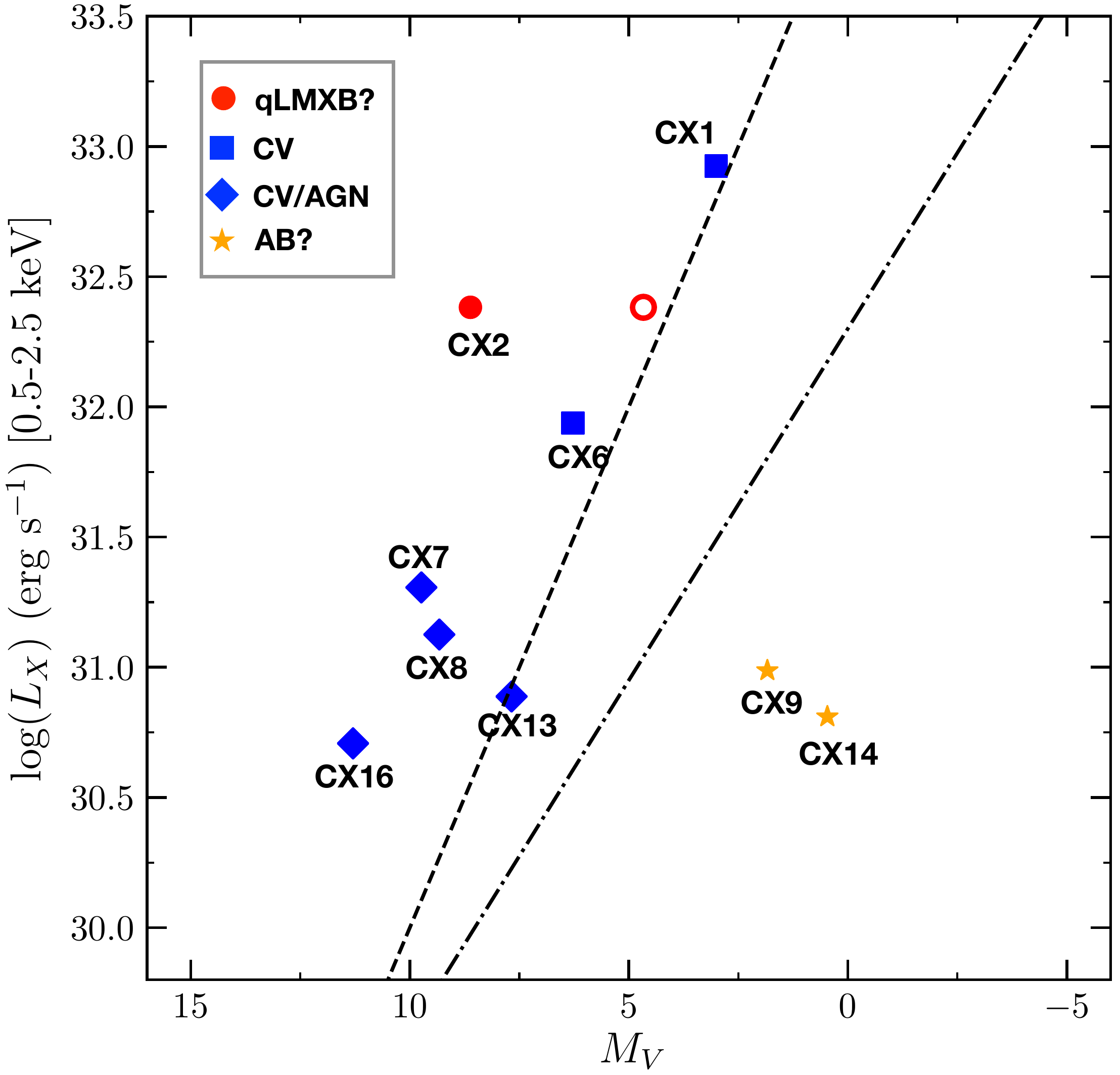}
    \caption{$0.5-2.5$ keV $L_X$ vs.\ absolute V band magnitudes (using $\mathrm{V_{555}}$ for CX7 and CX16, and  $\mathrm{V_{606}}$ for others). $L_X$ of CX1 is from \citet{stacey2011}. The dashed line corresponds to the $L_X = 34.0 - 0.4 M_V$ separatrix from \citet{Bassa04}, dividing cluster CVs and ABs. The dotted-dashed line corresponds to the $L_X = 32.3-0.27 M_V$ separatrix from \citet{Verbunt08}, marking the upper limit of $L_X$ for nearby ABs. CV candidates with confirmed cluster memberships are marked with filled blue squares. Possible CV/AGN candidates are marked with filled blue diamonds. The red filled circle marks the location of CX2 if we were to adopt the suggested counterpart. For comparison, the red open circle indicates the location of CX2 on the plot if we adopt the bright MS star mentioned in Sec. \ref{sec:cx2} as the counterpart.}
    \label{fig:xray_to_optical_ratio}
\end{figure}

We consider several possible natures for CX2; quiescent LMXB, MSP, CV, or AB in the cluster, or a background or foreground source. The NS atmosphere fit is consistent with a qLMXB nature, with a radius consistent with emission from a full NS surface. However, it is inconsistent with the expectation of emission from NS polar caps as seen in typical MSPs \citep{zavlin2002, Bogdanov06}.  Although the VMEKAL fit is statistically reasonable, the implied temperature is low. If CX2 is a member of M3, its X-ray luminosity of $2\times10^{32}$ erg/s is at the very high end of X-ray luminosities for globular cluster CVs, and all known CVs in globular clusters with $L_X > 10^{31}$ erg/s have much harder X-ray spectra \citep[e.g.][]{heinke2005, pooley2006}, so a CV nature can be ruled out empirically. \citet{Verbunt08} showed that nearby chromospherically active stars are limited in their X-ray luminosity, with log$L_X <32.3-0.27 M_V$, while active binaries in globular clusters are limited by log$L_X <34.0-0.4 M_V$ \citep{Bassa04}.  CX2's suggested counterpart has  $M_V=8.6$ (Tab.~\ref{tab:hst_counterparts}) and $L_X=2\times10^{32}$ erg/s (Tab.~\ref{tab:fitting_results}), so CX2 lies well above both limits (see Fig. \ref{fig:xray_to_optical_ratio}), strongly suggesting that it is not an AB. If CX2 were instead a foreground AB, the optical counterpart to CX2 should be brighter than $M_V=-0.3$; however, no bright cluster non-members were found within the error circle of CX2, arguing against a foreground system. 
%If CX2 is a cluster AB, the high $L_X$ would require an $ M_V\lesssim 4.0$ or $V \lesssim 19.1$. We found a
A bright cluster MS star ($P_\mu=98.1\%$) is located NW of the error circle (Fig.\ref{fig:finding_charts}), %likely to be the counterpart of a cluster AB, but, according to   the 2006 ACS/WFC observations, it -above the 19.1 limit
 but is still too faint (by 0.6 mags, $V\approx19.7\pm 0.03$; see Fig. \ref{fig:xray_to_optical_ratio}). Furthermore, X-ray bright cluster ABs generally have harder X-ray spectra than fainter ones (e.g. \citealt{heinke2005}), which contradicts  the soft nature of the source.  We therefore conclude that CX2 is most likely to be a qLMXB.

%Based on the high $L_X$ and soft nature of CX2, we conclude that it is most likely to be a quiescent LMXB in M3.
%A foreground active binary might be considered, but the ratio of X-ray luminosity to optical luminosity remains excessive for any distance. 

%Future deeper observations by instruments with larger collecting areas (e.g. {\it XMM-Newton}) can better distinguish between an NSATMOS v.s. a thermal plasma model (e.g. VMEKAL), and therefore between a qLMXB and other possibilities. 

 %{\color{red}--It would be good for us to estimate the chance likelihood of finding an optical star inside the error circle. For CX2, the relevant probability is the probability of finding *any* star in the error circle, since the star is nondescript. For other stars, it's worth computing the probability of finding a blue star in the error circle, or a red straggler, or a red giant.}
 
 Considering the relatively low central density and large mass of M3, it is interesting to consider whether this quiescent LMXB is more likely to be generated from a primordial binary (that is, via a similar evolutionary path as similar objects outside clusters), or via dynamical encounters (as the majority of  quiescent LMXBs and millisecond pulsars in GCs are thought to be produced).  To calculate the probability of this quiescent LMXB being a primordial binary, we use estimates of the total number of quiescent neutron star LMXBs in the Milky Way, which center around $10^3$ to $10^4$ systems \citep[][Heinke et al. 2018, in prep.]{Pfahl03,Kiel06,Jonker11,Britt14}. We use the Milky Way's stellar mass of $5\times10^{10}$ \Msun \citep{cox2000}, and an estimate of M3's stellar mass of $6\times10^5$ \Msun (adjusting the calculations of \citealt{Gnedin02}\footnote{http://www-personal.umich.edu/~ognedin/gc/vesc.dat} to use the average mass-to-light ratio of 1.86 from \citealt{Watkins15}), to predict 0.01 to 0.1 quiescent LMXBs in M3. Some population syntheses \citep[e.g.][]{Pfahl03} do generate up to $10^5$ LMXBs in the current Milky Way, which would predict of order 1 quiescent LMXB in M3. However, such a large number of quiescent LMXBs, in a Galaxy with of order 100 persistently bright LMXBs,  is strongly empirically disfavored by the observed ratio of $\sim$10-20 quiescent LMXBs per persistently bright LMXB in globular clusters \citep{heinke03d,heinke2005b}, and by the other empirical estimates cited above. From the dynamical side, 5 quiescent LMXBs are observed in 47 Tuc \citep{heinke2005b}, and M3 has a stellar interaction rate 19\% that of 47 Tuc \citep{Bahramian13}, so 0.97 dynamically-formed quiescent LMXBs are predicted in M3.  Thus, we find it 10-100 times more likely that this quiescent LMXB was formed dynamically, rather than primordially.
 
 As a point of interest, performing the same calculation for the quiescent LMXB in $\omega$ Centauri \citep{rutledge02a} indicates a prediction of 0.04-0.4 primordial systems, vs. 0.45 dynamically formed systems, suggesting that the $\omega$ Centauri quiescent LMXB has a decent chance to be a primordial system.

\subsection{CX6}
 A power-law fit to CX6's X-ray spectrum produces a negative photon index ($\Gamma = -0.7^{+0.8}_{-0.7}$) if we freeze the $n_H$ at the cluster value. However, $\Gamma$ becomes more physically reasonable when $N_H$ is allowed to float to a higher value. For example, at $N_H = 6.0\times 10^{22}~\mathrm{cm^{-2}}$, $\Gamma = 1.2^{+1.1}_{-1.3}$. %Although we cannot determine by how much the absorption should be increased, this does suggest that CX6 is a heavily absorbed source. 
 %{\color{red} Can we try to fit for $N_H$ and Gamma together?} (\green{With $N_H$ thawed, I still got a negative $\Gamma$, so, no.})

\begin{figure}
    \centering
    \includegraphics[scale=0.8]{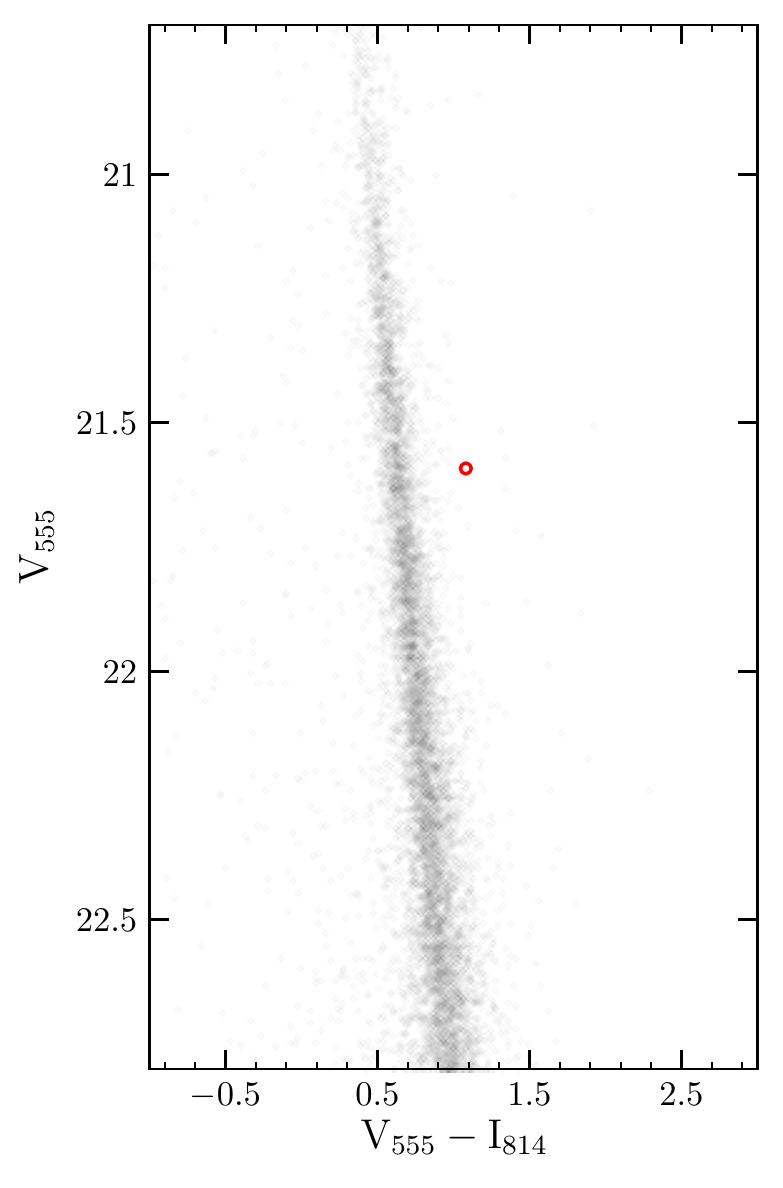}
    \caption{$\mathrm{V_{555}-I_{814}}$ CMD from the 2004 ACS/WFC observations. The red circle marks the location of the CX6 counterpart.}
    \label{fig:cx6_redexcess}
\end{figure}

 The optical counterpart to this source
 is very blue in UV colours ($E(UV-NUV) \approx -0.85$), but the $\mathrm{V-I}$ colour indicates a very large red excess in visible light, $E(V-I) \approx 0.37$, compared to the main sequence. This red excess was confirmed in the 2004 ACS/WFC $\mathrm{V_{555}-I_{814}}$ CMD (see Fig. \ref{fig:cx6_redexcess}) with $E(V-I)\approx 0.43$ with respect to the MS; however, using the $\mathrm{B_{435}}$ and $\mathrm{V_{555}}$ magnitudes from the same epoch, the star shows moderate blue excess relative to the MS (see Fig. \ref{fig:ksync_cmd}). %shows simultaneous blue and red excesses (with respect to the main sequence, $E(FUV-UV) \approx -0.85$ and $E(V-I) \approx 0.37$). 
 Assuming the same V magnitude as CX6, the $\mathrm{V-I}$ colour on the main sequence ($\mathrm{V-I} \approx 0.65$) corresponds to an F9V-G0V dwarf with $T_\mathrm{eff}\approx 5900~\mathrm{K}$, whereas CX6 corresponds to a K3V-K3.5V dwarf with $T_\mathrm{eff}\approx 4800~\mathrm{K}$ (spectral types and $T_\mathrm{eff}$'s are from \citealt{pecaut2013}). Again, if we apply the ``bloated envelope" scenario as for CX1, the change in $T_\mathrm{eff}$ requires that the companion to be bloated by a factor of $\sim 1.5$, which does not seem physically reasonable. 
 Although high measured $N_H$'s are common among AGNs, both CX6's PM and the associated $P_\mu$ strongly support its cluster membership. %With the numbers from
 In Sec. \ref{subsec:counterpart_search}, we estimated the number of chance coincidences of blue stars to be minuscule ($\approx 2.14\times 10^{-3}$), so the suggested counterpart is very unlikely to be spurious.
 %Another alternative is that CX6 is a very high inclination CV (explaining the hard X-ray colour; see e.g. \citealt{heinke2005} for the examples W8, W15, and AKO9), which shows eclipses (see e.g. \citealt{Edmonds03b}), and that the single V observation caught it in eclipse. --Oops, there are *two* V-I CMDs which *both* show it red, so this doesn't work.
 Therefore, the high $N_H$ might imply a CV seen edge-on, such as the CVs W8, W15, and AKO9 in 47 Tuc \citep{heinke2005}, but the red excess remains unexplained.
 Optical/UV variability analyses indicate strong UV and B variabilities to this star, suggestive of a CV nature, which might also be the cause of the observed simultaneous blue and red excess. Future spectroscopic study (e.g. with integral field units, such as MUSE) of this object could determine its nature. 
 %{\color{red} It would be good to generate multiple panels of Fig. 5, where we divide the stars by magnitude. I expect the fainter stars to have larger PM errors than brighter stars, so we can get a clearer picture if we subdivide the data into 3-4 magnitude bins.}

 %{\color{red} Perhaps we can check this by comparing the V, I mags from the ACS GC survey with V, I mags from the 2004 observations.}
 
\begin{figure*}
    \centering
    \includegraphics[scale=0.58]{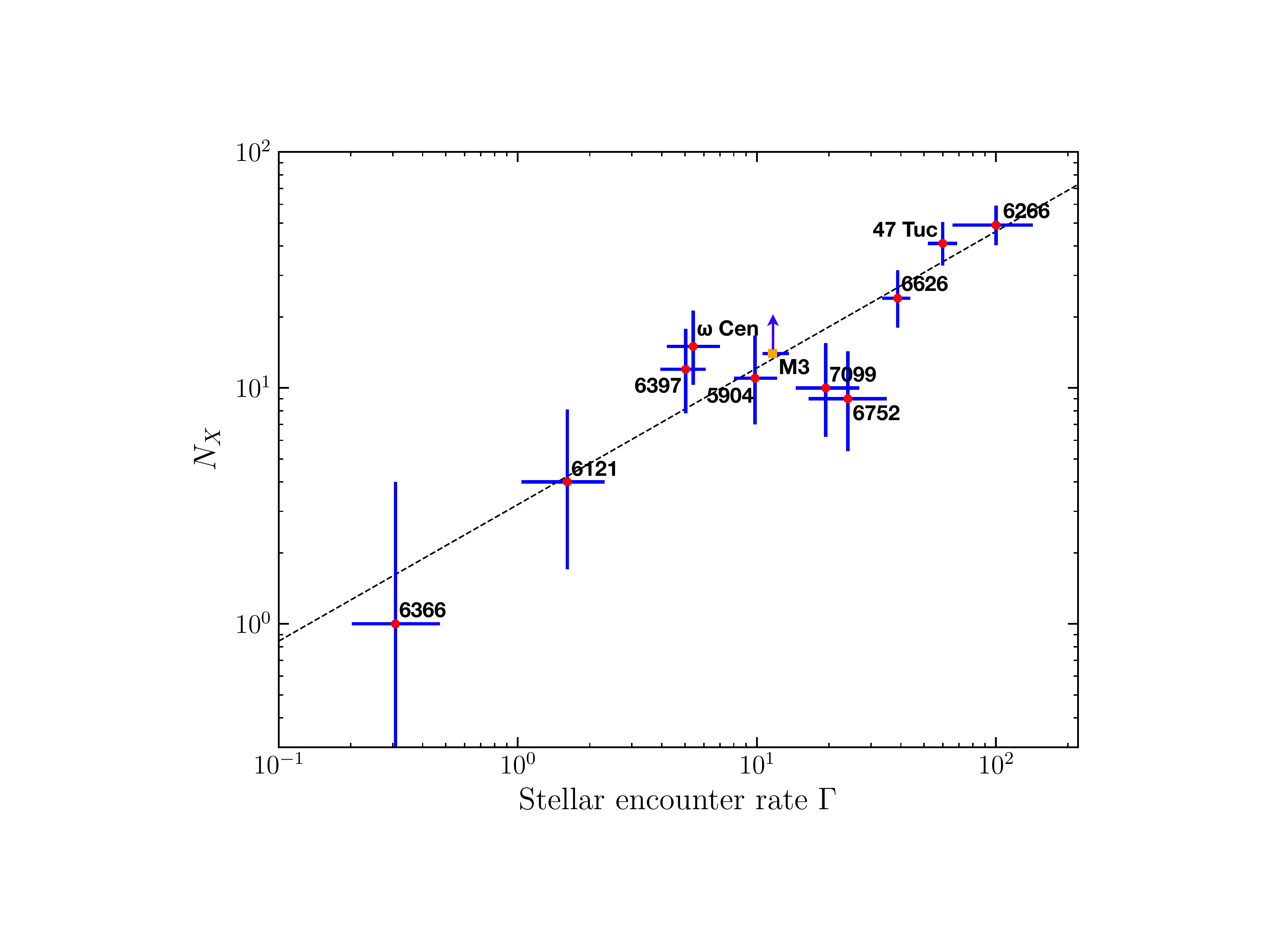}
    \caption{Number of non-AGN X-ray sources ($N_X$) in multiple GCs v.s. renormalised stellar encounter rates ($\Gamma$) from \protect\cite{Bahramian13}. $\Gamma$s have been renormalised such that NGC 6266 has a $\Gamma$ of 100. The lower limit inferred from our results of M3 is indicated with an orange square. The dashed line indicates the linear regression fit from \citet{lugger2017}.}
    \label{fig:nx_vs_gamma}
\end{figure*}

\subsection{CX8, CX12 and CX16}
 Similar to CX1 and CX6, the counterparts to these sources also show obvious UV excesses.  
 Considering the calculations of Sec. \ref{sec:chance_coincidence}, none of these counterparts are likely to be chance coincidences, so they should be CVs or background AGNs.
 The 2012 WFC3 observations only detected these three counterparts in the two UV bands ($\mathrm{UV_{275}}$ and $\mathrm{NUV_{336}}$). However, the 2004 ACS observations detected the counterpart to CX16, which shows a blue excess in the $\mathrm{B_{435}-V_{555}}$ CMD with a relatively larger error bar (see Fig. \ref{fig:ksync_cmd}). Although CX8 is apparently detected in the ACS GC survey catalogue (and therefore has a measured PM), this detection appears as a faint extension to a nearby bright star, which results in a larger photometric error and makes the blue excess more uncertain. Identifications of the faint optical counterparts to CX12 and CX16 are even more difficult, due to the lack of both accurate $\mathrm{V-I}$ colours, and proper motions.
 %{\color{red} Well, CX8 does have a measured proper motion--it's in Fig. 4. Slight rewrite?}
 %For instance, these UV sources could be cluster white dwarfs that are only chance coincidences with the X-ray error circles (this is of most concern for CX16, which lies on the apparent white dwarf sequence). 
 
 Another potential issue with these faint sources is that they have poorer localizations than bright sources. %which makes us check not only the stars within the error circles but also those nearby (e.g.
 The potential counterpart to CX16 is a faint UV source northeast of its error circle with a distance of $\approx 0.98\arcsec$ (or $\approx 1.84 P_\mathrm{err}$) from the nominal {\it Chandra} position, which  suggests that it may be a chance coincidence. 
 However, the {\it HST} lightcurve for CX16 reveals %weak 
 $\mathrm{NUV_{336}}$ variability on hours timescales (reduced $\chi^2 \approx 1.96$), suggesting a faint CV nature, since 
 %a single white dwarf should not be variable. 
 AGN tend to show less short-term variability.
 New HST imaging could reveal their PMs, testing an AGN hypothesis. %deeper optical images could reveal their SEDs, which should show a red optical component if they are indeed CVs. --unfortunately, many AGN also show a red optical component, due to the normal stellar population in the galaxy.

\subsection{CX7 and CX13}
Our suggested counterparts to CX7 and CX13 show moderately blue colors in $\mathrm{B_{435}-V_{555}}$ (2004) and/or $\mathrm{V_{606}-I_{814}}$ (2006) CMDs (see Fig. \ref{fig:ksync_cmd} and Tab. \ref{tab:hst_counterparts}). The counterpart to CX13 (within the $95\%$ error circle) was too faint to be detected in the 2012 WFC3 observations, so no proper motion is available for it. The counterpart to CX7 is outside the 95\% error circle ($\approx 0.9\arcsec$, or $\approx 1.7 P_\mathrm{err}$, from the nominal {\it Chandra} position). Because it was only covered by the 2004 ACS/WFC observations, only limited information can be drawn from the photometry. Considering the lack of PM measurements for these sources, we conclude that they could either be CVs or AGNs.%The large error in the I band magnitude makes the extent of the blue excess less certain.--it's in 2 CMDs at the same position, so the large error in one doesn't bother me much.

\subsection{CX9 and CX14}

 We found evolved stars that coincide with the error circles of CX9 and CX14. A red straggler (lying to the red of the giant branch; see \citealt{Geller17} for the definitions of red stragglers and sub-subgiants) and a subgiant are located in CX9's error circle, while a red giant is in CX14's error circle. The red straggler to CX9 was also detected in the Ksync photometry (2004 ACS/WFC observations; see Fig. \ref{fig:ksync_cmd}) to have an obvious red excess. CX9 and CX14 may be RS Canum Venaticorum (RS CVn) variables, which are systems with an evolved primary (e.g. F/K type subgiant or K type giant) in a close binary, where the primary has active chromospheric regions that induce large stellar spots. The observed X-ray emissions in RS CVn stars are thought to originate from active coronal regions on the primary and/or the secondary star.

 However, %based on the fact that this source is in the vicinity to the cluster center, 
 it is also possible that the evolved stars in these error circles are simply chance coincidences, or that the evolved stars have unseen compact companions. 
 %{\color{red} This is where we need quantitative estimates of the chance probabilities of finding a red giant, a subgiant, and a red straggler in an error circle.} 
 Considering %how much more unlikely that 
 the very low number of expected chance coincidences ($\approx 2.58\times 10^{-4}$) for %one of the few 
 a red straggler %in this field 
 to reside within the {\it Chandra} error circle, vs. a subgiant (with the number of chance coincidence $\approx 2.77\times 10^{-2}$),
 we regard the red straggler as a highly likely optical counterpart of CX9.  Red stragglers are rare, may be the product of mass transfer in a binary system, and have often been identified as X-ray sources \citep{Belloni98,Mathieu03,leiner2017}.
 %It is also possible that CX9 is an MSP based on its low luminosity ($L_{0.5-7.0}\approx 2.91\times 10^{31}~\mathrm{erg~s^{-1}}$) and proximity to the cluster center (its offset from the center of the core is $\sim 3.9\arcsec \approx 0.18 r_c$), where most MSPs were found (see e.g. \citealt[Figure 5]{camilo2005}). ---This seems like a weak argument.
 %Observationally, it is not unprecedented that an
 %Some MSPs have been associated with RS stars (e.g. PSR J1740-5340 in NGC6397, see \citealt{orosz2003}). ---Oops, actually a sub-subgiant, not RS.
 %{\it \color{red} I'm reluctant to go into this scenario in this much detail, as the evidence in favour of it for CX9 is quite weak. But we can discuss this together.}
 %Theories have also shown that RSs can be formed through mass transfer in close binaries (see \citealt{leiner2017}). For example, \cite{ivanova2017} show that an envelope-stripped subgiant can evolve to become an RS. 
 %The corresponding timescale on the RS regime for a subgiant with an initial mass of $1~M_\odot$ can be as long as $\sim 10^{8.5}~\mathrm{yr}$ when the envelope is stripped down to $M\sim 0.4~M_\odot$. RS is a viable type of companions to multiple XRB classes (e.g. LMXBs, MSPs, and etc.). 
 CX9 is %therefore 
 a very interesting target for future observations, and fortunately, it is not affected by serious crowding in the core, which makes it resolvable with instruments that have even larger PSFs.

\subsection{Estimate of XRBs and AGNs}
The number of AGN expected within the half-light radius can be estimated using the empirical model with three power-law components from \cite{Mateos2008}. Applying a soft ($0.5-2.0~\mathrm{keV}$) flux limit of $S = 5.22\times 10^{-16}~\mathrm{erg~s^{-1}~cm^{-2}}$, the model predicts that $N_\mathrm{AGN}(>S) \approx 2^{+3}_{-1}$ ($90\%$ confidence limits are from \citealt{gehrels1986}) within the half-light radius. Thus, we expect that 14$^{+1}_{-3}$ of our detected sources are likely members of M3, with CX12, and CX16 being plausible AGN (with no PM or $P_\mu$ available). CX8 does have a PM suggestive of a cluster member; however, further $P_\mu$ information is required for more secure identification. There are probably other AGNs among our X-ray sources without optical counterparts, which lie below our optical/UV detection limits. %{\color{red} Doesn't the PM of CX8 make it a less likely AGN? I agree CX12 is a good AGN candidate, and CX16 is also possible, though the variability argues against it.}

Three confirmed MSPs (and one candidate) are known in M3, two with timing solutions and thus known positions \citep{Hessels07}. The two known MSP positions do not correspond to any detected X-ray sources in our data. This is not surprising, since 
most MSPs observed in globular clusters have $L_X$(0.3-8 keV) between $10^{30}$--$10^{31}$ ergs s$^{-1}$ (e.g. only 1 of 23 MSPs with known positions in 47 Tuc has $L_X>10^{31}$ erg s$^{-1}$; \citealt{Bogdanov06,Ridolfi16,Bhattacharya17}).
%the typical X-ray luminosity range for MSPs is $L_X$(0.3-8 keV)=$2\times10^{30}$ to $10^{31}~\mathrm{erg~s^{-1}}$, which encompasses all but one (W) of the MSPs in 47 Tuc \citep{Bogdanov06}. 
Due to the larger distance to M3 (10.2 kpc; \citealt[2010 version]{harris1996}) and the relatively short exposures here, we do not have X-ray detections below $L_X\sim1.3\times10^{31}~\mathrm{erg~s^{-1}}$. However, it is still possible that one of our X-ray sources might be an MSP with an unusually high X-ray luminosity.

We compared the number of X-ray sources we found in M3 with those from other GCs, and with the expected numbers of dynamically formed X-ray binaries. 
We used stellar encounter rates ($\Gamma$) from \cite{Bahramian13}, and numbers of non-AGN X-ray sources ($N_X$) in multiple GCs from \cite{pooley2003} and \cite{lugger2007}. Since our luminosity limit ($L_{0.5-6.0}\approx 1.1 \times 10^{31}~\mathrm{erg~s^{-1}}$) is higher than Pooley's ($L_{0.5-6} = 4.0\times 10^{30}~\mathrm{erg~s^{-1}}$), the $N_X$ in M3 ($14^{+1}_{-3}$) reported here should be regarded as a lower limit. We compare our results of M3 with other GCs in Fig. \ref{fig:nx_vs_gamma} (where we have renormalised all $\Gamma$s so that $\Gamma$ for NGC 6266 is 100), together with a linear regression fit ($N_X \propto \Gamma^{0.58\pm 0.10}$ ) from \cite{lugger2017}. Deeper {\it Chandra} and {\it HST} observations would be helpful to verify the X-ray source content, and source classification, of M3. 
M3 will be a particularly helpful cluster, along with M13, M5, and $\omega$ Cen, in studying how lower-density clusters produce X-ray sources, as these clusters are likely to contain both primordial and dynamically-formed X-ray binaries.
%{\color{red} Perhaps we can use the formulation of Verbunt et al. 2008 to estimate the numbers of primordial and dynamically-formed XRBs we expect in this cluster (though, to a slightly lower Lx level).}
%determine a better localization for M3 on this plot.

\section{Conclusions}
\label{sec5}
Using $\sim 30~\mathrm{ks}$ of {\it Chandra} observations, we detected 16 X-ray point sources within the half-light radius of the globular cluster M3. The X-ray sources include the transient supersoft source and CV 1E1339, and a likely quiescent LMXB with a neutron star companion. Our optical/UV identification campaign has identified plausible optical and/or UV counterparts to 10 of 16 sources, including the previously identified 1E1339, a red straggler, a likely CV with unusually red optical colours, a faint red MS star to the qLMXB candidate, a possible giant (perhaps an RS CVn chromospherically active star), and five objects with UV and/or blue excesses, which may be CVs or AGNs.

\section*{Acknowledgements}

We thank J. Anderson for making his KSYNC code available for our use, and for discussions.
CH is supported by an NSERC Discovery Grant and a Discovery Accelerator Supplement.
Our X-ray analyses are based on data obtained from the {\it Chandra} Data Archive, observations made by the {\it Chandra} X-ray Observatory and published previously in cited articles. Optical analyses are based on observations made with the NASA/ESA {\it Hubble Space Telescope}, obtained from the data archive at the Space Telescope Science Institute. STScI is operated by the Association of Universities for Research in Astronomy, Inc. under NASA contract NAS 5-26555.
This work has made use of data from the European Space Agency (ESA) mission
{\it Gaia} (\url{https://www.cosmos.esa.int/gaia}), processed by the {\it Gaia}
Data Processing and Analysis Consortium (DPAC,
\url{https://www.cosmos.esa.int/web/gaia/dpac/consortium}). Funding for the DPAC
has been provided by national institutions, in particular the institutions
participating in the {\it Gaia} Multilateral Agreement.
This research has also made use of the NASA Astrophysics Data System (ADS) and software provided by the Chandra X-ray Center (CXC) in the application package CIAO.

\bibliographystyle{mnras}
\bibliography{m3_manuscript_cleaned} % if your bibtex file is called example.bib

\appendix

\section{Renormalization of SED models with $\chi^2$ Minimization method}

For a single component model, the $\chi^2$ is defined as 

\begin{equation}
    \chi^2 = \sum\limits_{i} \lrbs{\dfrac{\lrb{\alpha M_i - D_i}}{\sigma_i}}^2,
    \label{eq.a1}
\end{equation}

where $M_i$ and $D_i$ are the model value and the data value at $x_i$, respectively, and $\alpha$ is a normalization factor. We want to find $\alpha$ such that 

\begin{equation}
    \dfrac{\partial \chi^2}{\partial \alpha} = 0.
    \label{eq.a2}
\end{equation}

Plugging eq.(\ref{eq.a1}), one can solve for $\alpha$ and find that

\begin{equation}
    \alpha = \dfrac{\sum\limits_{i} \frac{M_i D_i}{\sigma_i^2}}{\sum\limits_{i} \frac{M_i^2}{\sigma_i^2}}.
    \label{eq.a3}
\end{equation}

For the case of an composite model with two additive components, i.e.

\begin{equation}
    M = \alpha M_1 + \beta M_2,
    \label{eq.a4}
\end{equation}

one can use similar method to find that

\begin{equation}
    \begin{split}
        &\alpha = -\dfrac{\lrb{\sum\limits_{i}\frac{M_{2,i}^2}{\sigma_i^2}}\lrb{\sum\limits_{i} \frac{D_i M_{1,i}}{\sigma_i^2}} - \lrb{\sum\limits_{i}\frac{M_{1,i}M_{2,i}}{\sigma_i^2}}\lrb{\sum\limits_{i} \frac{D_iM_{2,i}}{\sigma_i^2}}}{\lrb{\sum\limits_{i}\frac{M_{1,i} M_{2,i}}{\sigma_i^2}} - \lrb{\sum\limits_{i}\frac{M_{1,i}^2}{\sigma_i^2}}\lrb{\sum\limits_{i}\frac{M_{2,i}^2}{\sigma_i^2}}}\\
        &\beta  = -\dfrac{\lrb{\sum\limits_{i}\frac{M_{1,i}^2}{\sigma_i^2}}\lrb{\sum\limits_{i} \frac{D_i M_{2,i}}{\sigma_i^2}} - \lrb{\sum\limits_{i}\frac{M_{1,i}M_{2,i}}{\sigma_i^2}}\lrb{\sum\limits_{i} \frac{D_iM_{1,i}}{\sigma_i^2}}}{\lrb{\sum\limits_{i}\frac{M_{1,i} M_{2,i}}{\sigma_i^2}} - \lrb{\sum\limits_{i}\frac{M_{1,i}^2}{\sigma_i^2}}\lrb{\sum\limits_{i}\frac{M_{2,i}^2}{\sigma_i^2}}}
    \end{split}
    .
    \label{eq.a5}
\end{equation}
%%%%%%%%%%%%%%%%%%%%%%%%%%%%%%%%%%%%%%%%%%%%%%%%%%

% Don't change these lines
\bsp	% typesetting comment
\label{lastpage}

\end{document}